\documentclass[a4paper,11pt]{article}

\usepackage{latexsym}
\usepackage{amssymb}
\usepackage{graphicx}
\usepackage{amsmath,amsfonts}
\usepackage{bbm}

\usepackage{slashed}

\topskip \textwidth 480pt

\def\pa{\partial}
\def\*{\ast}

\def\ci{{\mathcal I}}

\def\s{\sigma}

\def\pa{\partial}

\def\be{\begin{equation}}
\def\ee{\end{equation}}
\def\bqn{\begin{eqnarray}}
\def\eqn{\end{eqnarray}}
\def\nn{\nonumber}

\def\theequation{\thesection.\arabic{equation}}

\newsavebox{\ver}
\newsavebox{\verp}

\newsavebox{\gorp}
\newsavebox{\toch}

\newcommand{\bee}{\begin{eqnarray}}
\newcommand{\eee}{\end{eqnarray}}

\newcommand{\rf}[1]{(\ref{#1})}
\def \foot {\footnote}
\def \no {\nonumber}
\def \la {\label}

\date{}
\begin{document}
\title{
{\small
\hfill Imperial-TP-AT-2012-05\\
\hfill MIFPA-12-40\\
\hfill NORDITA-2012-85\\{}}
\vskip 20pt
{\bf Semiclassical  
equivalence of   Green--Schwarz\\  and Pure--Spinor/Hybrid
formulations\\  of   superstrings    in  $AdS_5\times S^5$ and  $AdS_2\times S^2\times T^{6}$}
~\\
~\\
\author{Alessandra Cagnazzo$^{a}$, Dmitri Sorokin$^{b}$, Arkady A. Tseytlin$^{c,}$\footnote{Also at Lebedev Institute, Moscow.} ~and Linus Wulff$^d$
~\\
~\\
{\small $^a$\it Nordita, KTH Royal Institute of Technology and Stockholm University,}
~\\
{\small\it Roslagstullsbacken 23, SE-106 91 Stockholm, Sweden}
~\\
{\small $^b$ \it INFN, Sezione di Padova, via F. Marzolo 8, 35131 Padova, Italia}
~\\
{\small $^c$ \it Blackett Laboratory, Imperial College, London SW7 2AZ, U.K.}
~\\
\small{$^d$\it George P. \& Cynthia Woods Mitchell Institute}
{\small\it for Fundamental Physics and Astronomy,}
~\\
{\small\it Texas A\&M University, College Station, TX 77843, USA}
}
 }
\maketitle
\abstract
We  demonstrate  the equivalence   between   the worldsheet  one-loop   partition functions
computed near classical string solutions
in the Green--Schwarz and in the pure--spinor formulations of superstrings  in $AdS_5\times S^5$.
While their  bosonic sectors are the same in the conformal gauge,
their fermionic sectors  superficially appear to be  very different   (1st vs 2nd--derivative kinetic terms, presence vs absence of
 fermionic  gauge symmetry).
Still, we   show that the quadratic fluctuation  spectrum
of sixteen fermionic
 modes of the pure--spinor formulation is the same as in
 the Green--Schwarz superstring  and  the contribution of the extra ``massless'' fermionic modes
cancels against that of  the pure--spinor ghosts.
We also   provide evidence  for  a similar semiclassical equivalence  between   the Green-Schwarz  and the hybrid formulations
of superstrings in
   $AdS_2\times S^2\times T^6$    by studying several particular   examples of string solutions.

\thispagestyle{empty}
\newpage
\tableofcontents

\def \ci  {\cite} \def \la {\label}

\def \be {\begin{equation}}
\def \ee {\end{equation}}

\setcounter{page}{1}

\section{Introduction}
To describe superstring theory on $AdS \times M$ backgrounds with Ramond--Ramond fluxes one can use either the Green--Schwarz (GS) formulation \cite{Green:1983wt,Grisaru:1985fv,Metsaev:1998it}, or the pure--spinor  (PS)  formulation \cite{Berkovits:2000fe}, or, in some  low-dimensional cases,  a  hybrid model \cite{Berkovits:1999zq}.
The case of the $AdS_5\times S^5$ superstring has been studied in the GS and in the pure--spinor formulation
 (see, e.g.,  \cite{Arutyunov:2009ga,Beisert:2010jr,Berkovits:2000yr,Mazzucato:2011jt} for review).
 In the less supersymmetric backgrounds  $AdS_3\times S^3\times T^4$ and $AdS_2\times S^2\times T^6$
  the GS formulation (see  \cite{Babichenko:2009dk,Sorokin:2011rr} and references therein)
  and  hybrid models    \cite{Berkovits:1999zq,Berkovits:1999im,Berkovits:1999du} have been used.

  These   formulations are not on  an equal footing.  The GS  action has a clear physical origin (describing, e.g., the motion
  of a fundamental string soliton  in a type II supergravity background)  and can be defined in a reparametrization--invariant way
  (though for perturbative  quantization it requires a choice of a bosonic vacuum that spontaneously breaks some global  symmetries). The  PS  formulation is defined as a fermionic extension   of the  bosonic string in conformal gauge
   with second-derivative kinetic terms for the fermions and ghosts   added to ensure the BRST symmetry of the resulting 2d  conformal theory.
   While this construction is somewhat ad hoc  and the origin of the BRST invariance
      remains to be understood, it has  the  advantage that
   its perturbative   quantization does not require a choice of a bosonic vacuum and thus  can be, in principle,  performed   without breaking the underlying
   global symmetries.

The  general   relation   between the type II  GS superstring and the   PS  (or hybrid) formulation
 in  non--trivial backgrounds  remains an open problem.\footnote{For heterotic strings and in flat  space
 this  relation is, in fact,  understood much better (see e.g. \cite{Oda:2001zm,Matone:2002ft,Aisaka:2004ga,Berkovits:2004tw,Bandos:2012hp}).  For  generic curved superbackgrounds  a passage from the GS action to
 the pure--spinor one \cite{Berkovits:2001ue} was studied  in \cite{D'Auria:2008ny,Tonin:2010mm,Oda:2011nc}.}
Assuming it  exists, such a relation  is likely to   require   non--trivial (non--local) field redefinitions.
A way towards   understanding  how the equivalence   could  be established  is
  to study the  correspondence between the quantum
partition  functions   in  the two formulations  computed in the semiclassical expansion, i.e.
by expanding near a classical  bosonic  string  solution.


Almost all of the previous studies  of semiclassical strings in $AdS_5\times S^5$  and similar backgrounds was done using the GS formulation
(see, e.g., \ci{Drukker:2000ep,Frolov:2002av,Tseytlin:2003ii,Beisert:2010jr}).
This is not too surprising given that the structure of the fermionic part of the pure-spinor  action  is much more complicated
than that of the GS one. While  the quadratic  fermionic term in the GS action  is  roughly $\theta  D \theta \partial x $
(becoming $\theta  \tilde D \theta $  after a choice of  bosonic $x$-background  with  16  of 32  $\theta$'s decoupled due to
 kappa-symmetry) the  corresponding  term in the PS  action  is  $ D \theta  D \theta $
 which has  twice as many derivatives  and   none of the 32 $\theta$'s decouple  \emph{a priori}.

The study of the semiclassical expansion of the
 pure--spinor  superstring  was so far  done in the  BMN  limit
  \cite{Berkovits:2002zv},
  by expanding   near a rigid  circular 2-spin string in $S^3$ \cite{Roiban:2007}
   and, more recently, for  particular string  solutions in the  $R_t\times S^2$  part  of
   $AdS_5\times S^5$ \cite{Aisaka:2012ud}. The results were   suggesting the equivalence  (at least to 1-loop order  in the semiclassical expansion)
   with the corresponding  GS partition function.

In this paper  we generalise  the previous work of  \cite{Roiban:2007,Aisaka:2012ud}
and    explicitly prove the equivalence between the  GS and PS  worldsheet 1-loop   partition functions
computed by expanding near  a generic bosonic  string  solution in $AdS_5\times S^5$.
We show,  in particular,   that  16   fermionic modes of the pure--spinor formulation have the same
spectrum as the GS  fermions  while
all the other    modes decouple  and effectively cancel against the pure--spinor ghost contribution  in the partition function.

We also   argue for a similar   correspondence   between the GS and the hybrid    formulation
for the  $AdS_2\times S^2\times T^6$ superstring.
In this case  we do not provide a general proof but demonstrate the equivalence between the two fermionic sectors
in several special  cases. In particular, we  consider  the  background    of worldsheet instantons wrapping non--trivial cycles of $AdS_2\times S^2\times T^6$.

We  start in section 2 with a review  of the quadratic fermionic terms in the GS and in the PS actions expanded  near a bosonic
string solution.  We then prove the equivalence of their contributions to the one-loop  partition  function evaluated near a generic classical string solution
and  further   illustrate this  equivalence on the examples  of two   simple   infinite string solutions.

In  section 3   we    turn to the $AdS_2\times S^2\times T^6$     case and
discuss  the equivalence of the one--loop partition functions of the GS model  and the hybrid model for
 a general motion of the classical string in $AdS_2\times S^2$ and  also,  for   a few simple cases,
  when the string moves in $T^6$.

  Appendix A contains some  notation and conventions.
  In Appendix B we present a  comparison  between the
   GS and the hybrid model  quadratic fluctuation Lagrangians  in the case of a folded
   spinning  string in the $R_t\times S^2$  part of $AdS_2\times S^2$.

\setcounter{equation}0
\section{$AdS_5\times S^5$ Superstring}
In what follows we will be interested in one--loop partition functions  computed by expansion near
classical solutions, i.e. in the spectra of  quadratic fluctuation operators.
Since the pure spinor description of the superstring is \emph{a priori }  based on  the conformal gauge, we will use this
 gauge also in the Green--Schwarz  formulation.

The bosonic  part of the Lagrangian  which  is the same for the GS and  PS     formulations
is given by (we set the string tension to one)
\begin{equation}\label{ads5b}
\mathcal L_{\rm bose}= \frac12
(g^{_{AdS_5}}_{mn} \partial_ix^m\partial^ix^n+g^{_{S^5}}_{m'n'}\partial_iy^{m'}\partial^iy^{n'}),
\end{equation}
where $x^m$ are $AdS_5$ coordinates, $y^{m'}$ are  $S^5$ coordinates
 and $\xi^i=(\tau,\sigma)$ parametrize the worldsheet. We shall use the $AdS_5$ metric  in global coordinates  
\begin{equation}\label{gads5m}
ds^2=-\cosh^2\rho\ dt^2+d\rho^2+\sinh^2\rho\ (d\theta^2+\cos^2\theta d\phi_1^2+\sin^2\theta d\phi_2^2)\,.
\end{equation}
The non-zero components of the corresponding  $AdS_5$ spin connection are
\begin{eqnarray}
&&
\omega^{01}=-\sinh\rho\,dt\,,\qquad
\omega^{12}=\cosh\rho\,d\theta\,,\qquad
\omega^{13}=\cos\theta\cosh\rho\,d\phi_1\,,
\nonumber\\
&&
\omega^{14}=\sin\theta\cosh\rho\,d\phi_2\,,\qquad
\omega^{23}=-\sin\theta\,d\phi_1\,,\qquad
\omega^{24}=\cos\theta\,d\phi_2\,.
\label{ads5spinconn}
\end{eqnarray}
The Virasoro constraints on the bosonic fields are
\begin{equation}\label{virasoroads5}
G_{\tau\tau}+G_{\sigma\sigma}=0, \qquad G_{\tau\sigma}=0\,,\qquad \ \ \ \
G_{ij}\equiv g^{_{AdS_5}}_{ij}+g^{_{S^5}}_{ij}=e_i{}^Ae_j{}^B\eta_{AB} \ ,
\end{equation}
where  $e_i{}^A$ are the worldsheet pullbacks of the $AdS_5\times S^5$ vielbeins $\partial_i X^Me_M{}^A(X)$  ($A=a,a'$ and $X^M=x^m,y^{m'}$). These  constraints  express the fact  that
in the conformal gauge the induced metric is  conformally flat
\begin{equation}\label{confm}
G_{ij}=\phi(\xi)\,\eta_{ij}\,.
\end{equation}
The bosonic equations of motion in the conformal gauge are
\begin{equation}\label{beomads5}
\eta^{ij}\nabla_ie_j{}^A=0\,,
\end{equation}
where $\nabla_i$ is the worldsheet pullback of the $AdS_5\times S^5$ covariant derivative. Note  that the absence of torsion in the $AdS_5\times S^5$ connection  implies
 \begin{equation}\label{torsionless}
\nabla_{[i}e_{j]}{}^A=0\,.
\end{equation}

\subsection{Green--Schwarz formulation}
The quadratic  fermionic part of the GS Lagrangian on $AdS_5\times S^5$ in the conformal gauge is
(for our  notation and conventions regarding  spinors and  gamma--matrices see Appendix~A)
\begin{eqnarray}\label{GSads5}
 \mathcal{L}_{\rm GS}&=&i\Theta\left(\eta^{ij}-\varepsilon ^{ij}\,\sigma_3 \right)e_i{}^{A}\Gamma_{A}
 \,
 \left(\nabla_j+\frac{i}{2}\Gamma_{01234}\,\sigma_2\Gamma_{B} e_j{}^{B}\right)\Theta\,,
\end{eqnarray}
where the covariant derivative is given by $\nabla_i=\partial_i-\frac 14 \omega_i{}^{AB}\Gamma_{AB}$.\
 $\Theta^I$ $(I=1,3)$ are two 16--component Majorana--Weyl spinors of the same chirality and $\sigma_2^{IJ}$ and $\sigma_3^{IJ}$ are Pauli matrices.
 This  Lagrangian can be obtained from the complete $\mathbbm Z_4$--graded sigma--model action on the supercoset $\frac{PSU(2,2|4)}{SO(1,4)\times SO(5)}$ \cite{Metsaev:1998it}.\footnote{We use the indices $I=(1,3)$ instead of $I=(1,2)$ to indicate that $\Theta^I=(\Theta^1,\Theta^3)$ have, respectively, grading 1 and 3 with respect to the $\mathbbm Z_4$-automorphisms of the superisometry group $PSU(2,2|4)$ of the $AdS_5\times S^5$ superbackground. }

  The presence of the RR $F_5$ flux supporting the $AdS_5\times S^5$  space
  manifests itself in  the  $\Gamma_{01234}$-term  in the GS Lagrangian. To simplify the notation, let us define
\begin{equation}\label{G01234}
\gamma_*\equiv \Gamma_{01234},\qquad\qquad  \gamma^2_*=-1\,
\end{equation}
\begin{equation}\label{slashe}
\slashed e_i\equiv e_i{}^A\Gamma_A\,.
\end{equation}
Then the Lagrangian \rf{GSads5}  takes the form
\be\label{1GSads5}
 \mathcal{L}_{\rm GS}=i\Theta\left(\eta^{ij}-\varepsilon ^{ij}\,\sigma_3 \right)\slashed e_i
 \,
 \mathcal D_j\Theta\,,
\ee
where   
\be\label{kils}
\mathcal D_j=\nabla_j+\frac{i}{2}\gamma_*\,\sigma_2 \slashed e_j
\ee
is a generalized covariant derivative  appearing also  in the Killing--spinor  equation.

In terms of the  $\mathbbm Z_4$--graded
fermions  $\Theta^I=(\Theta^1,\Theta^3)$, i.e.
 $\Theta^1=\frac 12(1+\sigma^3)\Theta$ and $\Theta^3=\frac 12(1-\sigma^3)\Theta$,
  the Lagrangian \eqref{1GSads5}  may be explicitly written as
\be\label{GSads511}
 \mathcal{L}_{\rm GS}=-i\Theta^1\slashed e_-
 \,
\nabla_+\Theta^1 -i\Theta^3\slashed e_+
 \,
 \nabla_-\Theta^3-{i}\Theta^1\slashed e_-\gamma_*\slashed e_+\Theta^3\,,
\ee
where $\slashed e_\pm=\slashed e_\tau{}\pm \slashed e_\sigma$ and $\nabla_\pm=\nabla_\tau\pm\nabla_\sigma$.

The Virasoro constraints and the bosonic equations of motion satisfied by the classical string solutions, eqs. (\ref{virasoroads5}), (\ref{beomads5}) and (\ref{torsionless}), imply
\begin{equation}\label{gamma5pm}
(\slashed e_+)^2=0=(\slashed e_-)^2\,,\qquad \nabla_-\,\slashed e_+=0=\nabla_+\,\slashed e_-\,,
\end{equation}
\begin{equation}\label{ephi}
\slashed e_+{}\slashed e_-+\slashed e_- \slashed e_+=-4\phi(\xi)\,,
\end{equation}
where $\phi= \frac 12 \eta^{ij}e_i{}^Ae_j{}^B\eta_{AB}=-\frac12e_+{}^Ae_-{}^B\eta_{AB}$ is the conformal factor of the induced worldsheet metric \eqref{confm}.

Let us now introduce the two projectors\footnote{We assume here that $\phi\neq0$. This is not so for point--like string solutions which will be discussed further in what follows.}
\begin{equation}\label{P}
P_-=-\frac 1{4\phi(\xi)}\slash\!\!\! e_-\,\slash\!\!\! e_+\,,\qquad P_+=-\frac 1{4\phi(\xi)}\slash\!\!\! e_+\,\slash\!\!\! e_-\,
\end{equation}
satisfying,  in view of eqs. \eqref{gamma5pm}  and  \eqref{ephi}, the following relations
\begin{eqnarray}\label{PP}
P_++P_-=1\,,\qquad P_-P_-=P_-\,,\qquad P_+P_+=P_+\,,\qquad P_-P_+=0\,.\nonumber\\
\nabla_-P_-=-\nabla_-P_+\,,\qquad \nabla_+P_-=-\nabla_+P_+\,,\\
(\nabla_-P_-)P_+=P_-\nabla_-P_+=P_+\nabla_+P_-=(\nabla_+P_+)P_-=0.\nonumber
\end{eqnarray}
From these it also follows that
\begin{equation}\label{PP1}
\nabla_-P_-=(\nabla_-P_-)P_-,\quad \nabla_+P_-=P_-\nabla_+P_-\,,\quad \nabla_-P_+=P_+\nabla_-P_+,\quad \nabla_+P_+=(\nabla_+P_+)P_+\,.
\end{equation}
Using the above relations, the GS Lagrangian \eqref{GSads511} can be written as
\be\label{GSads51111}
 \mathcal{L}_{\rm GS}=-i\Theta_+^1\slashed e_-\,\nabla_+\Theta_+^1 -i\Theta_-^3\slashed e_+{}\,\nabla_-\Theta_-^3-{i}\Theta_+^1\slashed e_-\gamma_*\slashed e_+\Theta_-^3\,,
\ee
where
\begin{equation}\label{tpm}
\Theta^{1,3}_{\pm}=P_{\pm}\Theta^{1,3}\,.
\end{equation}
Thus  half of $\Theta^1$ and $\Theta^3$ drop out of the Lagrangian \eqref{GSads51111}.
This  is a consequence of the kappa--symmetry of the GS formulation. 
Indeed, the  GS action  is, in general,
 invariant under the off-shell  kappa--symmetry under which  not only the fermions but also the
bosonic string coordinates and the worldsheet metric are transforming.
However, when   expanding near a classical bosonic solution satisfying the Virasoro conditions
kappa--symmetry becomes equivalent simply to a shift  of fermions implying a degeneracy of the fermionic kinetic term
(i.e. decoupling of the gauge part of the fermions)
with the  corresponding projector being  ${\textstyle
\left(
\begin{array}{cc}
P_+&0\\
0&P_-	
\end{array}\right). } $

\subsection{Pure--spinor formulation}
The  quadratic  term  of the  $\Theta$--fermion part of the pure--spinor Lagrangian on $AdS_5\times S^5$ is
given by\footnote{For a detailed description of the PS  action  in the present context we refer the reader to \cite{Aisaka:2012ud}.}
\begin{eqnarray}\label{0fermpure}
 \mathcal{L}_{\rm PS}&=&i\Theta\left(\eta^{ij}-\varepsilon ^{ij}\,\sigma_3 \right)\slashed e_i
 \,
\mathcal D_j\Theta- 2\eta^{ij}\,\mathcal D_i\Theta\,\gamma_*\,\sigma_2\,\mathcal D_j\Theta\,,
\end{eqnarray}
where $\mathcal D_j$   was defined in \rf{kils}.
Note that the first term  of this Lagrangian is the same as the quadratic term in the
GS Lagrangian \eqref{GSads5}   while  the second term  is of  second order  in derivatives of
$\Theta$. The presence of this second  term breaks the kappa--symmetry  that was present in the GS  action.
Instead, the  PS action  contains  additional ghost   fields   and  is  required to be BRST invariant.
  The ghost sector consists of bosonic spinors $\lambda$, $\hat\lambda$,  $\omega$ and  $\hat\omega$ satisfying the pure spinor conditions
\begin{equation}\label{psc}
\lambda\Gamma^A\lambda=\hat\lambda\Gamma^A\hat\lambda=\omega\Gamma^A\omega=\hat\omega\Gamma^A\hat\omega=0\,,
\end{equation}
which  reduce the number
of independent components of each of the ghost fields   from 16 to 11.
The quadratic  part of the ghost Lagrangian is
\begin{equation}\label{psl}
\mathcal L_{\rm ghost}=\omega\nabla_-\lambda+\hat\omega\nabla_+\hat\lambda\,.
\end{equation}
 The   Lagrangian  \eqref{0fermpure} can be rewritten as follows
\begin{eqnarray}\label{fermpure}
 \mathcal{L}_{\rm PS}
=-i\varepsilon^{ij}\Theta \sigma_3 \slashed e_i
 \,
 \mathcal D_j\,\Theta
- 2\eta^{ij}\,\nabla_i\Theta\,\gamma_*\,\sigma_2\,\nabla_j\Theta+i\eta^{ij} \, \nabla_i\Theta \slashed e_j \Theta\, .
\end{eqnarray}
Another form, obtained by adding a total derivative, is
\begin{eqnarray}\label{fermpure1}
 \mathcal{L}_{\rm PS}
&=&- 2\,\nabla_i\Theta\,\gamma_*\,\sigma_2\,(\eta^{ij}+\varepsilon^{ij}\sigma_3)\mathcal D_j\Theta .\nonumber
\end{eqnarray}
In terms of the  $\mathbbm Z_4$--graded fermion  components
$\Theta=(\Theta^1,\Theta^3)$ this
 Lagrangian takes the following form
\begin{eqnarray}\label{fermpure2}
 \mathcal{L}_{\rm PS}
&=& 4i\,\nabla_-\Theta^1\,\gamma_*\nabla_+\Theta^3-i\Theta^1 \slashed e_+\nabla_-\Theta^1 - i\Theta^3 \slashed e_-\nabla_+\Theta^3.
\end{eqnarray}
To compare this Lagrangian with \rf{GSads51111} we split $\Theta^{1,3}$ entering \eqref{fermpure2} as in \eqref{tpm}
with the use of the projectors \eqref{P}--\eqref{PP1}  and get
\begin{eqnarray}
 \mathcal{L}_{\rm PS}
&=& 4i\,\nabla_-\Theta^1(P_++P_-)\,\gamma_*(P_++P_-)\nabla_+\Theta^3-
i\Theta_-^1 \slash\!\!\!e_+\nabla_-\Theta_-^1 - i\Theta_+^3 \slash\!\!\!e_-\nabla_+\Theta_+^3\nonumber\\
&&\nonumber\\
&=&4i\,\nabla_-\Theta_-^1\,P_+(\gamma_*-\gamma_*W_-^{-1}\gamma_*)P_+\nabla_+\Theta_+^3-i\Theta_-^1 \slash\!\!\!e_+\nabla_-\Theta_-^1 - i\Theta_+^3 \slash\!\!\!e_-\nabla_+\Theta_+^3\label{fermpure22}\\
&&+4i\,(\nabla_-\Theta_+^1+\nabla_-\Theta_-^1+\nabla_-\Theta_-^1\,P_+\gamma_* W_-^{-1})W_-(\nabla_+\Theta_-^3+\nabla_+\Theta_+^3+ W_-^{-1}\gamma_*P_+\nabla_+\Theta_+^3)\,,\nonumber
\end{eqnarray}
where
\begin{equation}\label{W}
W_-=P_-\gamma_*P_-
\end{equation}
and $W^{-1}$ is defined by
\be\label{W-}
 W_-W_-^{-1}=W^{-1}_-W_-=P_-\  .
\ee
Special cases in which $W_-^{-1}$ does not exist will be discussed in
 Section \ref{Dege}  below.
  It should be pointed out that the quadratic derivative terms in the first and the second line of \eqref{fermpure22}
  contain the matrix $\gamma_*$ which is sandwiched with different projectors $P_+$ and $P_-$, respectively. This is important for the proper separation of the terms containing $\Theta^1_+$ and $\Theta^3_-$ (second line) from the rest.

An important   feature  of this action is that
(the derivatives of)  $\Theta^1_+$ and $\Theta^3_-$ enter eq.  \eqref{fermpure22} only  linearly.
They can thus be integrated out producing a ``massless''  determinant of $\nabla_+\nabla_-$.
We will  then  be left with the first line in  \eqref{fermpure22} which should be compared with the GS action \eqref{GSads51111}. Indeed, if we substitute in \eqref{fermpure22} $X^1_+=\nabla_-\Theta_+^1P_-\equiv\nabla_-\Theta_+^1$ and $X^3_-=P_-\nabla_+\Theta_-^3\equiv\nabla_+\Theta_-^3$, we  find  that the equations of motion of $X^1_+$ and $X^3_-$ are $X^1_+=-(\nabla_-\Theta_-^1+\nabla_-\Theta_-^1\,P_+\gamma_* W_-^{-1})P_-$ and  $X^3_-=-P_-(\nabla_+\Theta_+^3+ W_-^{-1}\gamma_*P_+\nabla_+\Theta_+^3)$.   Then  the second line of \eqref{fermpure22} vanishes, and we are left with the first line containing only $\Theta^1_-$, $\Theta^3_+$, $\nabla_-\Theta_-^1P_+$ and $P_+\nabla_+\Theta_+^3$
 (the position  of the projectors $P_-$ and $P_+$ in the above relations is important).

The integration over  $\Theta^1_+$ and $\Theta^3_-$ in the second line of \eqref{fermpure22}
 requires extra care when the induced worldsheet geometry has a non--zero curvature.
 To have the equivalence with the GS formulation one should properly
 define the corresponding path integral measure (cf.  \cite{Schwarz:1992te}).
 We will  discuss  this  issue
  in more detail on the example of an
   infinite  string in $AdS_2$  in Section \ref{infads2}.

Let us now  comment on  the pure--spinor ghost sector \eqref{psl}.
Performing the following  transformations of the ghosts \cite{Aisaka:2012ud}
\be\label{ito}
\lambda\rightarrow V\lambda\,,\qquad \hat\lambda\rightarrow U \hat\lambda\,, \qquad\omega\rightarrow \omega\, V^{-1}\,,\qquad
\hat\omega\rightarrow\hat\omega\, U^{-1}\,
\ee
such that\footnote{One should not confuse these transformations with local Lorentz rotations. Using the  latter one
would not be able to completely remove the non--trivial spin connection from the derivatives. Note that being effectively one dimensional, eqs. \eqref{UV} can always be solved for V and U.}
\begin{equation}\label{UV}
V^{-1}\partial_-V=\frac 14\omega_-{}^{AB}\Gamma_{AB},\qquad U^{-1}\partial_+U=\frac 14 \omega_+{}^{AB}\Gamma_{AB}\,,
\end{equation}
the kinetic terms \eqref{psl} for the transformed pure spinors will contain the trivial
 partial derivatives $\partial_\pm$ only:
\begin{equation}\label{psl1}
\mathcal L_{\rm ghost}=\omega\partial_-\lambda+\hat\omega\partial_+\hat\lambda\,,
\end{equation}
so that their contribution
to the 1-loop partition function will be given simply by
 massless flat space Laplace  determinants.

Note that analogous transformations of $\Theta^1$ and $\Theta^3$ can be used \cite{Aisaka:2012ud} to convert the covariant derivatives of the fermionic Lagrangians \eqref{fermpure2}, \eqref{fermpure22} and \eqref{generic} into
simple  partial derivatives
\begin{equation}\label{trans}
\Theta^1\rightarrow V \Theta^1\,,\qquad\qquad \Theta^3 \rightarrow U\Theta^3. 
\end{equation}
Such  transformations may be useful
for simplifying the fermionic Lagrangian when  considering  particular
 examples of  string solutions, but  in the generic case the analysis of the Lagrangian in the form \eqref{fermpure2}
   turns out to be technically simpler.

\subsection{Relation between Green--Schwarz and pure--spinor formulations}

Below  we  will show the equivalence between the  1-loop partition functions in  the
GS and PS  formulations computed  for generic   string solutions in $AdS_5\times S^5$.
It is useful to start with a
 simpler case in which the  string moves only    in    $AdS_5$.

\def \no {\nonumber}

\subsubsection{String motion in $AdS_5$ }

In the special case when only $AdS_5$ (or, by analytic continuation, only $S^5$)   string coordinates are non-zero
 the projectors $ \slashed e_{\pm}$ and $P_\pm$ (anti)commute with $\gamma_*=\Gamma_{01234}$ and the Lagrangian \eqref{fermpure22} simplifies to\foot{The   sign of  the first term  can be  changed by changing
the sign of  $\Theta^1$. Here $\phi$ is the conformal factor from \rf{ephi}.}
\begin{eqnarray}
 \mathcal{L}_{\rm PS}&=&
 4i\,\nabla_-\Theta_-^1\,P_+\gamma_*P_+\nabla_+\Theta_+^3-i\Theta_-^1 \slashed e_+\nabla_-\Theta_-^1
 - i\Theta_+^3 \slashed e_-\nabla_+\Theta_+^3\nonumber\\
&&\ \ +4i(\nabla_-\Theta_+^1+\nabla_-\Theta_-^1)P_-\gamma_*P_-(\nabla_+\Theta_-^3+\nabla_+\Theta_+^3)\,\nonumber\\
&=&-{i}{\phi}^{-1}\,\nabla_-\Theta_-^1\,\slash\!\!\!e_+\gamma_*\slash\!\!\!e_-\nabla_+\Theta_+^3-i\Theta_-^1 \slash\!\!\!e_+\nabla_-\Theta_-^1 - i\Theta_+^3 \slash\!\!\!e_-\nabla_+\Theta_+^3\no \\
&&-{i}{\phi}^{-1}(\nabla_-\Theta_+^1+\nabla_-\Theta_-^1)\slash\!\!\!e_-\gamma_* \slash\!\!\!e_+(\nabla_+\Theta_-^3+\nabla_+\Theta_+^3)\,. \label{GSads5m}
\end{eqnarray}
Upon integrating out $\Theta^1_+$ and $\Theta^{3}_-$, we are left with the
 effective Lagrangian given  by the first line of \eqref{GSads5m};  we will denote it as
  ${\mathcal L}_1$. To compare   it
  with the GS Lagrangian \eqref{GSads51111}, let us rewrite  ${\mathcal L}_1$
   in the following
   first order form by introducing the Lagrange multiplier spinors  $\Psi^1$ and $\Psi^3$
   (i.e. integrating out $\Psi^1$ and $\Psi^3$  leads back to the first line of \eqref{GSads5m})
\begin{eqnarray}\label{1GSads5m}
 \mathcal{L}_1&=&-i\,\Psi_+^3\,\slash\!\!\!e_-\gamma_*\slash\!\!\! e_+\Psi_-^1+2i\Psi_-^1\slash\!\!\!e_+\nabla_-\Theta_-^1+2i\Psi_+^3\slash\!\!\!e_-\nabla_+\Theta_+^3\nonumber\\ &&-i\Theta_-^1 \slash\!\!\!e_+\nabla_-\Theta_-^1 - i\Theta_+^3 \slash\!\!\!e_-\nabla_+\Theta_+^3\nonumber\\
 &=&i\Psi_-^1 \slash\!\!\!e_+\nabla_-\Psi_-^1 +i\Psi_+^3 \slash\!\!\!e_-\nabla_+\Psi_+^3+i\,\Psi_-^1\,\slash\!\!\!e_+\gamma_*\slash\!\!\! e_-\Psi_+^3 \nonumber\\ &&-i\tilde\Theta_-^1\slash\!\!\!e_+\nabla_-\tilde\Theta_-^1 - i\tilde\Theta_+^3 \slash\!\!\!e_-\nabla_+\tilde\Theta_+^3\,,
 \end{eqnarray}
 where we introduced
 \be \tilde\Theta_-^1=\Theta^1_--\Psi_-^1\ ,\qquad \qquad \tilde\Theta_+^3=\Theta^3_+-\Psi_+^3\,.  \ee
From  \eqref{1GSads5m} we  conclude
 that the action for sixteen independent fermions $\Psi_-^{1}$ and $\Psi^3_+$ is  the same as for
 the GS fermions in \eqref{GSads51111},\footnote{Note that in the corresponding GS
 action  the $\nabla_+$ derivative acts on $\Theta^1$ and  $\nabla_-$ acts on $\Theta^3$, i.e. relating
  the GS and the PS  actions requires   exchanging  $\Theta^1$ with  $\Theta^3$   \cite{Aisaka:2012ud}.
As we have already mentioned, the fermionic part of the PS action   can be formally
 obtained (see eq. \eqref{fermpure}) by
 adding to the GS  action  an extra  second--derivative  term for the fermions.
 It then happens (as we shall also see on some  examples) that the resulting model
  can be related to
   the GS superstring which has the opposite sign of the Wess--Zumino term compared to the original one we started with.
   This change of the WZ term sign
   is equivalent to  interchanging  
   $\Theta^1$ and $\Theta^3$.}
 while the sixteen $\tilde\Theta^1_-$ and $\tilde\Theta^3_+$ modes   decouple and   contribute just ``massless''
 determinants  to the partition function.

 We thus conclude that when a classical string worldsheet is embedded  only  in $AdS_5$  (or only in $S^5$)
 the fermionic sector of the PS action  produces  the same one-loop
  contribution as the  GS  fermions   up to additional
   ``massless'' determinants. The latter   should
    be canceled  by the pure-spinor ghost  contributions  as required  by a consistent
    count of degrees of freedom  (and as a necessary requirement for having a consistent  flat-space limit).

\subsubsection{Generic  string motion}

 In the case of a generic classical string motion in $AdS_5\times S^5$ the first line of the Lagrangian \eqref{fermpure22} can also be put into  first order form analogous to \eqref{1GSads5m}
 \begin{eqnarray}\label{generic}
 \mathcal{L}_1=i\,\Psi^3_+\,M^{-1}\Psi^1_-+2i\Psi^1_-\slash\!\!\!e_+\nabla_-\Theta_-^1+2i\Psi^3_+\slash\!\!\!e_-\nabla_+\Theta_+^3-i\Theta_-^1 \slash\!\!\!e_+\nabla_-\Theta_-^1 - i\Theta_+^3 \slash\!\!\!e_-\nabla_+\Theta_+^3\nonumber\\
 =i\Psi^1_- \slash\!\!\!e_+\nabla_-\Psi^1_- +i\Psi^3_+ \slash\!\!\!e_-\nabla_+\Psi^3_++i\,\Psi^3_+\,M^{-1}\Psi^1_- 
   -i\tilde\Theta_-^1 \slash\!\!\!e_+\nabla_-\tilde\Theta_-^1 - i\tilde\Theta_+^3 \slash\!\!\!e_-\nabla_+\tilde\Theta_+^3
\ , \   \end{eqnarray}
 where
\begin{equation}\label{X}
M= \frac{1}{4\phi}\,\slash\!\!\!e_-(\gamma_*-\gamma_* W_-^{-1}\gamma_*) \slash\!\!\!e_+\,,\ \ \ \
\qquad MM^{-1}=\slash\!\!\!e_-\slash\!\!\!e_+\,.
\end{equation}
To establish the relation with the GS fermionic Lagrangian \eqref{GSads51111} it remains to show that the matrix $M^{-1}$ coincides with $\slashed e_-\gamma_*\slashed e_+$. For a class of solutions  describing
 strings moving  in the $R_t\times S^2$  part of $AdS_5\times S^5$ this was  demonstrated in \cite{Aisaka:2012ud}.
We shall now prove that this is so for any generic classical  string solution.

We should first find $W_-^{-1}$ of eq. \eqref{W-}. To this end let us analyse the structure of the matrix $W_-=P_-\gamma_*P_-$. We split $\slashed e_{\pm}$ as follows
\begin{equation}
\slashed e_{\pm}=\slashed a_\pm+ \slashed s_\pm\ , \qquad
\ \ \  \ \slashed a_\pm\equiv e_\pm{}^a(x) \Gamma_a\,, \qquad  \slashed s_\pm\equiv e_\pm{}^{a'}(y) \Gamma_{a'}
\end{equation}
where  $\slashed a_\pm$ and $\slashed s_\pm$  are, respectively, the pullbacks
of the gamma--contracted $AdS_5$ and $S^5$ vielbeins satisfying
\begin{eqnarray}\label{as}
 &\{\slashed a_\pm,\slashed s_\pm\}=0,\qquad [\slashed a_\pm,\gamma_*]=0\,,\qquad \{\slashed s_\pm,\gamma_*\}=0,&\nonumber\\
&\slashed a_+\,\slashed a_+\,=-\slashed s_+\,\slashed s_+\,,\qquad \slashed a_-\,\slashed a_-\,=-\slashed s_-\,\slashed s_-\,.&
\end{eqnarray}
The relations in the last line are due to the Virasoro constraints \eqref{gamma5pm}. Using these relations we find that
\begin{eqnarray}\label{pgp}
&W_-=P_-\gamma_*P_-=\frac 1{16\phi^2} \slash\!\!\! e_{-}\slash\!\!\! e_{+}\gamma_*\slash\!\!\! e_{-}\slash\!\!\! e_{+}=
\frac 1{16\phi^2}e_{-}\gamma_*(\slash\!\!\! a_{+}-\slash\!\!\! s_{+})\slash\!\!\! e_{-}\slash\!\!\! e_{+}&\nonumber\\
&=-\frac 1{16\phi^2} \slash\!\!\! e_{-}\gamma_*(\slash\!\!\! a_{+}-\slash\!\!\! s_{+})\,\slash\!\!\!\! R= -\frac 1{16\phi^2} \slash\!\!\! e_{-}\slash\!\!\! e_{+}\gamma_*\,\slash\!\!\!\! R&\nonumber\\
&=\frac 1{4\phi} P_-\gamma_*\,\slash\!\!\!\! R=\frac 1{4\phi}\,\slash\!\!\!\! R\,\gamma_*P_-\,,&
\end{eqnarray}
where
\begin{equation}\label{R}
\slash\!\!\!\! R=\slash\!\!\! a_+\slash\!\!\! a_--\slash\!\!\! a_-\slash\!\!\! a_++\slash\!\!\! s_+\slash\!\!\! s_--\slash\!\!\! s_-\slash\!\!\! s_+\,,\qquad [\slash\!\!\!\! R,\gamma_*]=0\,.
\end{equation}
Note that $\slashed R_{AdS_5}=\slashed a_+\slashed a_--\slashed a_-\slashed a_+\,$ and $\slashed R_{S^5}=-(\slashed s_+\slashed s_--\slashed s_-\slashed s_+)$ are the worldsheet pullbacks of the $AdS_5$ and $S^5$ curvature.

We also find that
\begin{equation}\label{+g}
\slash\!\!\! e_+\slash\!\!\! e_-\gamma_*\slash\!\!\! e_+=\slash\!\!\!\! R\,\gamma_*\slash\!\!\! e_+\,.
\end{equation}
For a generic classical string solution the matrix $\slash\!\!\!\! R$ is invertible (special cases for which this is not true are discussed in the next section).
Then
\begin{equation}\label{w-}
W_-^{-1}=-{4\phi}\, \slash\!\!\!\! R^{-1}\gamma_*P_-\,=-{4\phi}\,P_-\gamma_*\,\slash\!\!\!\! R^{-1}.
\end{equation}
Substituting eq. \eqref{w-} into \eqref{X} we get
\begin{equation}\label{M}
M= \frac{1}{4\phi}\,\slash\!\!\!e_-(\gamma_*-\gamma_* W_-^{-1}\gamma_*) \slash\!\!\!e_+= \frac{1}{4\phi}\,(\slash\!\!\!e_-\gamma_*\slash\!\!\!e_+-4\phi\,\slash\!\!\!e_-\gamma_*P_-\slash\!\!\!\! R^{-1}\slash\!\!\!e_+).
\end{equation}
Its `inverse' matrix is $M^{-1}=\slash\!\!\!e_-\gamma_*\slash\!\!\!e_+$. Indeed, using eq. \eqref{+g}, one checks that
\begin{eqnarray}\label{m-m}
M\,\slash\!\!\!e_-\gamma_*\slash\!\!\!e_+=
\frac{1}{4\phi}\,(\slash\!\!\!e_-\gamma_*\slash\!\!\!e_+-4\phi\,\slash\!\!\!e_-\gamma_*P_-
\slash\!\!\!\! R^{-1}\slash\!\!\!e_+)\,\slash\!\!\!e_-\gamma_*\slash\!\!\!e_+ \nonumber\\
=
 -\slash\!\!\!e_-\gamma_*P_+\gamma_*\slash\!\!\!e_+-\slash\!\!\!e_-\gamma_*P_-
 \slash\!\!\!\! R^{-1}\slash\!\!\!e_+\slash\!\!\!e_-\gamma_*\slash\!\!\!e_+
=-\slash\!\!\!e_-\gamma_*P_+\gamma_*\slash\!\!\!e_+-\slash\!\!\!e_-\gamma_*P_-\gamma_*\slash\!\!\!e_+
=\slash\!\!\!e_-\slash\!\!\!e_+\,.
\end{eqnarray}
We have thus proved that the ``mass matrix''
 $M^{-1}$ which appears in the first--order form of the PS fermionic Lagrangian \eqref{generic} coincides with that of the GS Lagrangian \eqref{GSads51111}. This shows that around any
 classical string solution with non--degenerate matrix $\slashed R$ in \eqref{R} sixteen fermions of the pure--spinor formulation  have the same fluctuation spectrum
  as  the GS fermions  while the rest of the PS fermions  are effectively ``massless''.
\\

To summarise,  the  correspondence between the   one-loop partition functions in the GS and PS formulations
can be   described as follows.
 The contribution of the 10 bosonic fluctuation modes  is of course the same in the two cases.
In the GS formulation the 8 (pairs of) physical fermionic modes
contribute the determinant of  a  Dirac-like  operator  
and there is also a trivial (flat-space)  conformal ghost determinant\footnote{Here  we are assuming that
we    use  flat rather than  an  induced 2d metric to define the determinants
in the conformal gauge, see \ci{Drukker:2000ep}  for a related discussion.}
$(\det\partial^2)^2$.
In the PS formulation  the above analysis implies  that  8 (pairs of)
 fermionic modes contribute  the same determinant as the GS fermions
  while  the remaining  24 fermionic modes
     produce massless determinants $(\det\partial^2)^{16+8}=(\det\partial^2)^{24}$.
     The factor $(\det\partial^2)^{16}$ comes from the integration of $\Theta^1_+$ and $\Theta^3_-$ in the second line of \eqref{fermpure22} while $(\det\partial^2)^8$  originates
     from the massless modes in \eqref{generic}.
     In the PS formulation there are no conformal ghosts but
     there are 22 chiral and 22 anti--chiral pure spinor ghosts\footnote{To integrate the pure--spinor ghosts contributing into the path integral with the Lagrangian \eqref{psl}, one should take into account the constraints \eqref{psc}. At the expense of $D=10$ covariance,  each of the constraints can be solved explicitly in terms of eleven independent parameters. } which contribute $(\det\partial^2)^{-22}$.
     In total, we are left with the ``massless''  determinant   factor
      $(\det\partial^2)^2$  which is the same
       as the contribution of the conformal ghosts in the GS formulation.

       We thus conclude that the one--loop partition functions in the two formulations do
       match near a generic classical string solution. 

\subsubsection{Degenerate cases \label{Dege}}

There are two special  classes of solutions which are not covered by the above analysis:

(i) Solutions with degenerate   2d
induced metric for which the  conformal factor vanishes, $\phi=0$. This happens for
 point--like (\emph{e.g.} BMN)  solutions.
 In these cases we cannot directly define the projection operators in (\ref{P}),
 and the analysis should be done, for instance, by first performing a suitable re--scaling of fermions with $\phi$ in the corresponding non--degenerate case and only then taking the limit $\phi\rightarrow 0$.
  The BMN limit of the $AdS_5\times S^5$ superstring in the
  pure spinor formulation was studied in \cite{Berkovits:2002zv} and was shown to be equivalent to that of the GS superstring. In Section \ref{BMNlimit}
  we will show that a similar  equivalence  holds also   between  the GS
  and the hybrid  formulation of the $AdS_2 \times S^2 \times T^6$ superstring
  with a  BMN  geodesic running along $S^2$ and $T^6$.

(ii)  Solutions for which $\slashed R$ in (\ref{R}) is not invertible.
Note that the fact that $\slashed R(\slashed R_{AdS_5}+\slashed R_{S^5})\propto \eta^{ij}(g^{AdS_5}_{ij}-g^{S^5}_{ij})$ means that $\slashed R$ is  invertible
unless $\eta^{ij}g^{AdS_5}_{ij}=\eta^{ij}g^{S^5}_{ij}=\phi$. Whether $\slashed R$ is
 invertible for solutions with $\eta^{ij}g^{AdS_5}_{ij}=\eta^{ij}g^{S^5}_{ij}=\phi$  should be checked case by case.

An example of a rigid rotating string for which $\slashed R$ is not invertible is the $J\rightarrow 0$ limit of the
circular   string solution of \ci{Frolov:2003qc} (see
eqs.  (3.12)-(3.13) in  \cite{Tseytlin:2010jv})
 for which $\slashed R=\slashed R_{AdS_5}=\slashed R_{S^5}=0$.
 As this is  a limit of a solution for which $\slashed R$ is invertible,
  we expect that the PS and GS formulations
  should   still  agree also in this limit.
  Let us  now show this explicitly. In this case
\begin{equation}\label{eg}
\slashed e_\pm=\slashed a_\tau\pm\slashed s_\sigma\quad \Rightarrow \quad \slashed e_+\gamma_* =\gamma_* \slashed e_-\qquad \Rightarrow\qquad \slashed e_+\gamma_*\slashed e_-=\slashed e_-\gamma_*\slashed e_+=P_\pm\gamma_*P_\pm=0\,.
\end{equation}
Since  here the induced  metric is flat   we have
$\nabla_\pm=\partial_\pm$
and, due to \eqref{gamma5pm} and \eqref{eg},
\begin{equation}\label{de}
\partial_+\slashed e_+=-\partial_+(\gamma_*\slashed e_-\gamma_*)=0\,,\quad  \partial_-\slashed e_-=-\partial_-(\gamma_*\slashed e_+\gamma_*)=0\quad \Rightarrow \quad \partial_\pm P_\pm=\partial_\mp P_\pm=0\,.
\end{equation}
As a result, the GS Lagrangian   takes the simple ``massless''  form
\begin{eqnarray}\label{0GSads51111}
 \mathcal{L}_{\rm GS}&=&-i\Theta_+^1\slashed e_-\,\partial_+\Theta_+^1 -i\Theta_-^3\slashed e_+{}\,\partial_-\Theta_-^3
\ , \end{eqnarray}
while  the PS Lagrangian  becomes
\begin{eqnarray}\label{singular}
\mathcal{L}_{\rm PS}
&=&4i\,\partial_-\Theta_-^1 \gamma_*\partial_+\Theta^3_-+4i\,\partial_-\Theta_+^1\,\gamma_*\partial_+\Theta_+^3
-i\Theta_-^1 \slash\!\!\!e_+\partial_-\Theta_-^1
-i\Theta_+^3 \slash\!\!\!e_-\partial_+\Theta_+^3\,.\nonumber
\end{eqnarray}
From the form of this Lagrangian in which  the (derivatives of) $\Theta^1_+$ and $\Theta^3_-$ enter only linearly
 it is clear that the integration over  the fermionic fields will only contribute
 to the partition function with ``massless'' determinants:  One finds indeed a factor
 $\sim (\det \partial^2)^{8+24}$ times the contribution $(\det \partial^2)^{-22}$ of the pure--spinor ghosts,
  which  is  the same as in  the GS case.

The matrix $\slashed R$ may, in principle, be degenerate also in (certain limits of) more complicated cases like elliptic strings or non--rigid strings. Then the number of ``massive''
 fermionic modes in both the Green--Schwarz and the pure--spinor formulation will
 be determined by the rank of the matrix \eqref{+g}, i.e. $\slashed e_-\slashed R\gamma_*\slashed e_+$. The analysis
 will  follow  along the similar  lines, by   combining
  the consideration made in  the above singular case and in
  the  case of non--degenerate   invertible part of $\slashed e_-\slashed R\gamma_*\slashed e_+$.

Below we will illustrate  the general proof of  the semiclassical equivalence of GS and PS formulations
given above by explicitly  verifying  this equivalence
for   two   simple examples of   string solutions.

\subsection{Examples of  string  solutions in $AdS_5\times S^5$ \la{exx}}

We shall     consider two  simple  limits of the folded spinning string
moving in the $AdS_3$ part of $AdS_5$  \ci{Gubser:2002tv,Frolov:2002av}. 
For generic values of parameters the corresponding  string coordinates in the $AdS_5$ metric (\ref{gads5m}) are
\begin{equation}\label{pulsating}
t=\kappa\tau,\qquad \phi_2=\omega\tau,\qquad \rho=\rho(\sigma)\,,\qquad \phi_1=0 \ , \ \ \ \ \ \ \
 \theta=\frac\pi 2\,,
\end{equation}
where
\begin{equation}\label{vrhopuls}
\rho'^2=\kappa^2\cosh^2\rho-\omega^2\sinh^2\rho\,,\qquad \rho{''}=(\kappa^2-\omega^2)\sinh\rho\cosh\rho\,.
\end{equation}
One special case  \ci{Frolov:2002av}
   is  when $\omega =0$  with  periodicity constraint in $\sigma$ removed --
  one finds then an open-string  solution  that  represents  an infinite string in $AdS_2$
    stretched all the way to the boundary with
  \be \la{rhoads2}
t=\kappa\tau,\qquad \qquad   \tanh\frac{\rho}{2} = \tanh {\frac{\kappa \sigma}{2}}   \ , \ \ \ \
\rho'= \kappa \cosh \rho \ .
  \ee
 The corresponding induced  metric is the $AdS_2$ one with  the curvature $R^{(2)}=-2$.\foot{Written in Poincar\'e coordinates
 this  Minkowski solution  is equivalent to a worldsheet ending on an  infinite straight line at the   boundary.
 This  is 1/2 BPS  configuration; fluctuations near it were studied in \ci{Drukker:2000ep}.}

Another  limiting case is  that   of an infinite--spin  long folded string when
   $\kappa \to \infty$  so that
   \begin{equation}\label{ls}
t=\phi_2 = \kappa\tau,\qquad \
\qquad \rho=\kappa\sigma\,,   \qquad \qquad  \ \  \phi_1=0\ , \ \ \ \ \ \ \ \  \theta=\frac{\pi }{2} \ .
\end{equation}
   Here   the string   also  reaches the boundary,
    but   having a   spin,  it has to be embedded at least into $AdS_3$.\foot{After a Euclidean worldsheet continuation and conformal   transformation the corresponding Poincar\'e patch solution is a null cusp surface
    \ci{Kruczenski:2007cy} .}


\subsubsection{Infinite  string in  $AdS_2\subset AdS_5$\label{infads2}}

Choosing $\kappa=1$  in \rf{rhoads2} the corresponding
 $AdS_2$ metric and spin connection are
\begin{eqnarray}\label{globalads2}
ds^2&=&-\cosh^2\rho  dt^2+d\rho^2\,, \qquad e^0=\cosh\rho dt\,, \qquad e^1=d\rho\,,\nonumber\\
\omega^{ab}&=&-\sinh\rho\ dt\, \varepsilon^{ab}\,,\qquad \nabla_\tau=\partial_\tau+\frac 12\sinh\rho\Gamma_{01}\,,\qquad \nabla_\sigma=\partial_\sigma\,.
\end{eqnarray}
Defining the $AdS_5 \times S^5$  fluctuation fields  as
\be
t=\tau+\frac{\tilde t(\tau,\sigma)}{\cosh \rho}\,,\qquad \rho=\rho(\sigma)+\tilde \rho(\tau,\sigma)\,, \qquad \tilde \theta(\tau,\sigma)\,,\qquad \tilde \phi_{1,2}(\tau,\sigma)\,,\qquad \tilde y^{m'}(\tau,\sigma)\,,
\ee
we  can write  the bosonic part
of the  quadratic fluctuation  Lagrangian as\foot{The bosonic as well as GS fermionic
fluctuation Lagrangians can be found  from the general folded  string  conformal gauge
expressions in \ci{Frolov:2002av}  by setting $\omega=0$ therein.}
\bee\label{ads5ads5b}
&\mathcal L_{\rm bose}=-\frac{1}{2}((\nabla\tilde t)^2 +\cosh^2\rho\,\tilde t^2)+ \frac{1}{2}((\nabla\tilde\rho)^2+\cosh^2\rho\,\tilde\rho^2)+\frac{1}{2}((\pa\tilde\theta)^2+2\cosh^2\rho\,\tilde\theta^2)&\nonumber\\
&+\frac{1}{2}((\pa\tilde \phi_1)^2+2\cosh^2\rho\,\tilde \phi_1^2)+\frac{1}{2}((\pa\tilde \phi_2)^2+2\cosh^2\rho\,\tilde\phi_2^2)+\frac{1}{2}(\pa \tilde y^{m'})^2\,,&
\eee
where $\nabla_i=\partial_i-\omega_i$ is the conventional $AdS_2$ covariant derivative with $\omega_i$ defined in \eqref{globalads2}. We see that for this solution there are two bosonic modes of (non--constant) ``mass" $\cosh\rho$ and three modes of ``mass" $\sqrt2\cosh\rho$, with $\rho$ being the classical solution $\rho(\sigma)$. The sum of the squared masses being
\begin{equation}\label{mb}
\sum m^2_b=(2+3\times 2)\cosh^2\rho=8\cosh^2\rho\ .
\end{equation}
Using  (\ref{rhoads2}) and \eqref{globalads2} in the definition of the projectors (\ref{P}) we find
\begin{equation}
\slashed e_\pm=\cosh\rho\,\Gamma_\pm\,,\qquad
\Gamma_\pm=\Gamma_0\pm\Gamma_1\ , \qquad P_\pm=\frac{1}{2}(1\pm\Gamma_{01})\,.
\end{equation}
 The   GS fermionic Lagrangian \eqref{GSads51111} then reduces to
\be\label{AdS2ads5final}
\mathcal{L}_{\rm GS}
= -i\cosh\rho\left(\Theta_+^1\Gamma_-\,\partial_+\Theta_+^1
+\Theta_-^3\Gamma_+\,\partial_-\Theta_-^3
+4\cosh\rho\,\Theta_+^1\Gamma_{234}\Theta_-^3\right)\,.
\ee
Note that the terms with the $AdS_2$ spin connection $\omega^{01}\Gamma_{01}$ vanish in \eqref{AdS2ads5final}, since $\Gamma_{\pm}\Gamma_{01}=\mp\Gamma_{\pm}$ is a symmetric matrix. For the same reason the overall $\cosh\rho$ factor can be removed by re--scaling $\Theta^I$.
Here $\Theta^1_+=P_+\Theta^1$ and $\Theta^3_-=P_-\Theta^3$ are $8+8$ fermionic modes
 which represent  8 physical degrees of freedom of ``mass" $\cosh\rho$.
 One can check that the UV   divergences (proportional to sum of mass-squared terms)   cancel between
 the  bosonic modes (see \eqref{mb})  and the   fermionic modes  in \rf{AdS2ads5final}. We are effectively assuming   that  fluctuation operators are defined   with respect to flat
 fiducial metric rather than the curved induced one so the conformal ghost contribution is trivial, cf. the discussion in
 \ci{Drukker:2000ep}.

In the pure--spinor  action written for the present solution
the fermions $\Theta_+^1$ and $\Theta_-^3$ completely decouple from the rest,
and the Lagrangian  \eqref{GSads5m}, in which we replace  the
first line with its first--order counterpart \eqref{1GSads5m}, takes the following form
\begin{eqnarray}\label{fermpure0}
{\mathcal L}_{\rm PS}&=&
i\cosh\rho\left(
\Psi_-^1\Gamma_+\,\partial_-\Psi_-^1
+\Psi_+^3\Gamma_-\,\partial_+\Psi_+^3
-4\cosh\rho\,\Psi_-^1\Gamma_{234}\Psi_+^3
\right)
\nonumber\\
&&{}
-i\cosh\rho
\left(
\tilde\Theta_-^1\Gamma_+\,\partial_-\tilde\Theta_-^1
+\tilde\Theta_+^3\Gamma_-\,\partial_+\tilde     \Theta_+^3
\right)
-4i\nabla_-\Theta_+^1\Gamma_{234}\nabla_+\Theta_-^3\,.
\end{eqnarray}
Here the  first line is equivalent to the GS Lagrangian \eqref{AdS2ads5final} and thus
 produces the same contribution to the partition function, while the
 fermions  $\tilde \Theta_-^1$   and    $\tilde \Theta_+^3$   in the   second line are obviously massless.

To analyse the contribution of the remaining
$\Theta^{1}_+$ and $\Theta^{3}_-$ fermions into the partition function, let us perform the transformation
  \eqref{trans}  with  $\omega_\pm=-2\sinh\rho\,\Gamma_{01}$ so that
$$
e^{\frac 14\int d\xi^+\omega_+}=e^{\frac 14\int d\xi^-\omega_-}=e^{-\frac 12\ln\cosh\rho \Gamma_{01}}\,.
$$
Then the last term in  \eqref{fermpure0} takes the form
\begin{equation}\label{fermpure01}
{\mathcal L}' =  -4i\nabla_-\Theta_+^1\Gamma_{234}\nabla_+\Theta_-^3  \ \to \ \   -{4i}\,{\cosh\rho}\,\partial_-\Theta_+^1\Gamma_{234}\partial_+\Theta_-^3.
\end{equation}

The integration over $\Theta_+^1$   and $\Theta_-^3$  (done, e.g.,
by first  changing the variables  to  $\partial_-\Theta_+^1 \to   Y_1, \
\partial_+\Theta_-^3\to   Y_3$ and assuming the local  contribution of the integral over $Y_1, Y_3$ is canceled against
a factor in the path integral measure, cf. \ci{Schwarz:1992te})
produces a massless determinant $(-\partial_+\partial_-)^{16}= (\partial^2)^{16}$.
Then the total fermionic  contribution to the partition function in the  PS model  \eqref{fermpure0}  is  $
 [\det (\partial^2-\cosh^2\rho)]^8\,(\det\partial^2)^{24}\,,
$
which matches the  corresponding   GS  result  (extra  massless determinants
are compensated by the ghosts as discussed above).\footnote{The total  result for
the partition function of the infinite straight string   surface should   be trivial as discussed in
 \ci{Drukker:2000ep}.}

Let us elaborate on the point  that the treatment of  the integral over $\Theta_+^1$ and $\Theta_-^3$
depends on assumptions   about the definition of the corresponding path integral. \emph{I.e.} the statement of equivalence between the GS and PS models   assumes  a particular prescription for the measure.
Observing that  in the present   case of curved induced geometry
  $\nabla_+\nabla_--\nabla_-\nabla_+=\cosh^2\rho\,\Gamma_{01}
$
and integrating   by parts in the last term in  \eqref{fermpure0}  or in \rf{fermpure01}
we may rewrite it as
\begin{equation}\label{fermpure012}
{\mathcal L}'=-{4i}\nabla_-\Theta_+^1\Gamma_{234}\nabla_+\Theta_-^3=-{4i}\nabla_+\Theta_+^1\Gamma_{234}\nabla_-\Theta_-^3
+4i\cosh^2\rho\ \Theta_+^1\Gamma_{234}\Theta_-^3\,.
\end{equation}
 Introducing the Lagrange multipliers $\Lambda_+^1$ and $\Lambda_-^3$
 we may convert  this into 1-st order form as
\begin{eqnarray}\label{1stfermpure012}
{\mathcal L}''&=&i\cosh\rho\left(-2\Lambda_+^1\Gamma_-\nabla_+\Theta_+^1
-2\Lambda_-^3\Gamma_+\nabla_-\Theta_-^3
+4\cosh\rho\,\Lambda_+^1\Gamma_{234}\Lambda_-^3
+4\cosh\rho\,\Theta_+^1\Gamma_{234}\Theta_-^3\right)\nonumber\\
&&=
i\cosh\rho(Y_+^1\Gamma_-\partial_+Y_+^1
+Y_-^3\Gamma_+\partial_-Y_-^3
+4\cosh\rho\,Y_+^1\Gamma_{234}Y_-^3)
\nonumber\\
&&{}
-i\cosh\rho(X_+^1\Gamma_-\partial_+X_+^1
+X_-^3\Gamma_+\partial_-X_-^3
-4\cosh\rho\,X_+^1\Gamma_{234}X_-^3)\,,
\end{eqnarray}
where $X^{1,3}_\pm=\frac 1{\sqrt{2}}(\Lambda^{1,3}_\pm+\Theta^{1,3}_\pm)$ and $Y^{1,3}_\pm=\frac 1{\sqrt{2}}(\Lambda^{1,3}_\pm-\Theta^{1,3}_\pm)$.
Since the last two lines  here  look like  copies of the GS fermionic Lagrangian \eqref{AdS2ads5final}
one might  naively  conclude that  (taking also into account  the first line of \eqref{fermpure0})
 the PS  Lagrangian 
  has three times more ``massive'' fermionic modes than the GS one.
  However, this   conclusion is premature as it depends  on the assumption that
  the introduction of the  auxiliary fields  in \rf{1stfermpure012}  did not produce  additional   determinant   factors.
  This is  not so in general as  they enter into  \rf{1stfermpure012}  with non-trivial   background-dependent
  factors, so the final result depends on a proper definition of path integral measure.


\subsubsection{Long spinning string in $AdS_3\subset AdS_5$\label{lss} }

Let us now consider the long infinite--spin   string solution \rf{ls}.
The non-zero components of the pull-backs of the vielbeins to the worldsheet here are
\begin{eqnarray}\label{e}
e_\tau{}^0=\kappa\cosh\rho\,,\qquad e_\tau{}^4=\kappa\sinh\rho\,,\qquad e_\sigma{}^1=\kappa\,,\\
e_\tau{}^A\Gamma_A=\kappa(\cosh\rho\Gamma_0+\sinh\rho\Gamma_4),\,\qquad e_\sigma{}^A\Gamma_A=\kappa\Gamma_1\,,\nonumber
\end{eqnarray}
and  the induced metric is  flat $ds^2=\kappa^2(-d\tau^2+d\sigma^2)$.
The Lagrangian for the
(appropriately redefined) bosonic fluctuations around this solution
(\cite{Frolov:2002av}) has the following form
\begin{eqnarray}\label{bopulsads5}
\mathcal L_{\rm bose}= &-\frac{1}{2}(\partial\tilde t)^2+ \frac{1}{2}(\partial\tilde\rho)^2+ \frac 12(\partial\tilde\phi_2)^2 +4\kappa\,\tilde t\partial_\sigma\tilde\phi_2-4\kappa\,\tilde\rho\partial_\tau\tilde\phi_2&\\
&+\frac{1}{2}((\pa\tilde\theta)^2+2\kappa^2\tilde\theta^2)+\frac{1}{2}((\pa\tilde \phi_1)^2+2\kappa^2\tilde \phi_1^2)+\frac{1}{2}(\pa \tilde y^{m'})^2\,.\nn&
\end{eqnarray}
It effectively describes two modes of mass $\sqrt{2}\kappa$, one mode of mass $2\kappa$ and seven massless modes. Their mass squared sum is $8\kappa^2$.

To  write down the GS fermionic action  let us note that
the  relevant components  of the spin connection (\ref{ads5spinconn})  are
\begin{equation}
\omega^{01}=-\sinh\rho\,dt
\,,\qquad\omega^{14}=\sin\theta\cosh\rho\,d\phi_2
\,,\qquad\omega^{24}=\cos\theta\,d\phi_2
\,,
\end{equation}
i.e., since $\theta=\frac \pi 2$,    we get
 $-\frac{1}{4}\omega_\tau{}^{AB}\Gamma_{AB}=\frac{\kappa}{2}(\sinh\rho\ \Gamma_0+\cosh\rho\ \Gamma_4)\Gamma_1$.
Thus one can perform the Lorentz rotation 
\begin{eqnarray}\label{Lr}
\Theta\,\rightarrow\, e^{\frac\rho 2\Gamma_{04}}\Theta, \qquad e^{-\frac\rho 2\Gamma_{04}}(\sinh\rho\Gamma_0+\cosh\rho\Gamma_4)e^{\frac\rho 2\Gamma_{04}}\,\rightarrow \,\Gamma_4\ ,  \nonumber \\
\ \ \   e^{-\frac\rho 2\Gamma_{04}}(\sinh\rho\Gamma_4+\cosh\rho\Gamma_0)e^{\frac\rho 2\Gamma_{04}}\,\rightarrow \,\Gamma_0\,,
\end{eqnarray}
under which the fermion  covariant derivative becomes
\begin{equation}\label{cd}
\nabla_\tau=\partial_\tau+\frac\kappa 2\Gamma_{41}\,,\qquad \nabla_\sigma=\partial_\sigma+\frac\kappa 2\Gamma_{04}\,,\qquad \slash\!\!\!\!\nabla=\eta^{ij}e_i{}^A\Gamma_A\nabla_j=\kappa(-\Gamma_0\partial_\tau+\Gamma_1\partial_\sigma)=\kappa\slash\!\!\!\partial\,.
\end{equation}
Using   \eqref{Lr} and \eqref{cd}
the corresponding  fermionic part of the
  GS Lagrangian \eqref{1GSads5}  takes  the following form
  \begin{equation}\label{gspuls}
\mathcal{L}_{\rm GS}
=2i\,( \Theta_+\slash\!\!\!\partial\, \Theta_++\kappa \Theta_+\Gamma_{234}\sigma_1 \Theta_+)\,,
\end{equation}
where $ \Theta_+=\frac{1}{2}(1+\Gamma_{01}\sigma_3) \Theta$ are 16 component fermions which carry 8 physical degrees of freedom of mass $\kappa$. Their mass squared sum  exactly cancels the mass squared sum of the bosonic modes \eqref{bopulsads5}.

 After  performing the Lorentz rotation \eqref{Lr} and \eqref{cd}, the quadratic fermionic part of the PS
 Lagrangian \eqref{fermpure} takes the form\footnote{Because of the
 simplicity of the form of the PS Lagrangian in this  background, here we do not need to
 split $\Theta$ into $\Theta_\pm^1$ and $\Theta^3_\pm$  as in  \eqref{fermpure2}.
 }
\begin{eqnarray}\label{fermpure4}
 \mathcal{L}_{\rm PS}
&=&
-2\eta^{ij}\,\nabla_i\Theta\,\Gamma_{01234}\,\sigma_2\,\nabla_j\Theta-{i\kappa^2}\Theta\Gamma_{234}\,\sigma_1\,\Theta
-i\kappa\Theta(1+\Gamma_{01}\sigma_3)\,\slash\!\!\!\partial\Theta\,.
\end{eqnarray}
 Substituting the explicit expressions \eqref{cd} for $\nabla_i$ and integrating by parts we get
 \if{}
 \begin{eqnarray}\label{fermpure04}
 \mathcal{L}_{\rm PS}
&=&
-2\eta^{ij}\,\partial_i\Theta\,\Gamma_{01234}\,\sigma_2\,\partial_j\Theta
-{i\kappa^2}\Theta\Gamma_{234}\,\sigma_1(1-\Gamma_{01}\sigma_3)\,\Theta
\nonumber \\
&&
-i\kappa\Theta(1+\Gamma_{01}\sigma_3)\,\slash\!\!\!\partial\Theta\, -2\kappa\Theta\Gamma_{023}\sigma_{2}\partial_\tau\Theta-2\kappa\Theta\Gamma_{123}\sigma_{2}\partial_\sigma\Theta\,.
\end{eqnarray}
\fi
\begin{eqnarray}\label{fermpure004}
 \mathcal{L}_{\rm PS}
&=&
2\Theta\,\Gamma_{01234}\,\sigma_2\,\partial_i\partial^i\Theta
-{i\kappa^2}\Theta\Gamma_{234}\,\sigma_1(1-\Gamma_{01}\sigma_3)\,\Theta
\nonumber \\
&&
-i\kappa\Theta(1+\Gamma_{01}\sigma_3)\,\slash\!\!\!\partial\Theta\, -2\kappa\Theta\Gamma_{023}\sigma_{2}\partial_\tau\Theta-2\kappa\Theta\Gamma_{123}\sigma_{2}\partial_\sigma\Theta\,.
\end{eqnarray}
The direct computation of the    determinant of the corresponding second--order kinetic operator  $D_{\rm PS}$
gives
\begin{equation}\label{det}
\det D_{\rm PS}=\left[\det (\pa^2-\kappa^2)\right]^{8}\,(\pa^2)^{24} \ .
\end{equation}
Thus  the pure spinor  fermion  massive contribution is  the same   (8
 fermionic modes of mass $\kappa$)  as the GS  one.



\setcounter{equation}0
\section{$AdS_2\times S^2 \times T^6$ superstring}

There are several  $AdS_2\times S^2 \times T^6$ backgrounds in type IIA and IIB
string theories that are supported by RR fluxes (see \cite{Sorokin:2011rr} for a review and references).
They preserve only 8 of 32 supersymmetries and are invariant under the superisometry group $PSU(1,1|2)$.
The dynamics of a superstring whose motion is restricted to the four--dimensional subspace $AdS_2\times S^2$
can  be described by two different $\frac{PSU(1,1|2)}{SO(1,1)\times U(1)}$
supercoset sigma--models with 8 fermionic fields. The first one  is of the Green--Schwarz type \cite{Zhou:1999sm},
having  similar structure to the $AdS_5\times S^5$   GS superstring \cite{Metsaev:1998it}, and
 the second one is an $N=(2,2)$ worldsheet superconformal $\frac{PSU(1,1|2)}{SO(1,1)\times U(1)}$
 sigma--model \cite{Berkovits:1999zq}
 which is the $AdS_2\times S^2$ counterpart of the supercoset sector of the $AdS_5\times S^5$ pure--spinor
 model.

When extended to the whole $AdS_2\times S^2 \times T^6$ space-time,
 the GS superstring sigma--model
 gets enlarged with  6 bosons   and
  24 fermions  which couple in a non--trivial
   way to the $4d$ $\frac{PSU(1,1|2)}{SO(1,1)\times U(1)}$  supercoset sector
    \cite{Sorokin:2011rr}. On the other hand, the $\frac{PSU(1,1|2)}{SO(1,1)\times U(1)}$
    sigma--model of \cite{Berkovits:1999zq}
    gets enlarged with an $N=(2,2)$ worldsheet superconformal  Ramond-Neveu-Schwarz   $T^6$--sector
   which couples to the supercoset sector only indirectly  via the (super)Virasoro constraints.
   The latter  model is called hybrid \cite{Berkovits:1996bf,Berkovits:1999zq} since it is constructed  using both the
   target--space spinors  as in the GS formulation    and the worldsheet spinors as in the RNS formulation.

The purely bosonic sectors of the GS superstring and the hybrid model  for  $AdS_2\times S^2 \times T^6$    are the same
 and in the conformal gauge have the form
\begin{equation}\label{bose}
{\mathcal L}_{bose}=\frac{1}{2} \eta^{ij}(g^{_{AdS_2}}_{ij}+g^{_{S^2}}_{ij}+g^{_{T^6}}_{ij})=\frac{1}{2} \eta^{ij}G_{ij}(X)\,,
\end{equation}
where $G_{ij}(X)$  ($X^M=(x^m,x^{\hat m},y^{m'})$)
is the induced worldsheet metric, $g^{_{AdS_2}}_{ij}=g^{_{AdS_2}}_{mn}(x)\partial_i x^m\partial_jx^n$, $g^{_{S^2}}_{ij}=g^{_{S^2}}_{\hat m\hat n}(\hat x)\partial_i x^{\hat m}\partial_jx^{\hat n}$ and $g^{_{T^6}}_{ij}=g^{_{T^6}}_{a'b'}\partial_i y^{a'}\partial_jy^{b'}$.

The Virasoro constraints
 and  the bosonic equations of motion are as in \eqref{virasoroads5} and \eqref{beomads5} in which $e_j{}^A$ are now the worldsheet pullbacks of the $AdS_2\times S^2\times T^6$ vielbeins $d\xi^i\partial_iX^Me_M{}^A(X)$ ($A=a,\hat a,a'$).

 The fermionic sectors of the two models are
  significantly different and to find the direct relation between them, which may involve
   a non-trivial (non--linear and non--local)  change of variables is   an open  problem.
   In what follows we shall study the relation between the two formulations of the $AdS_2\times S^2 \times T^6$ superstring by comparing
   their  quadratic fermionic actions  (and thus  one-loop  semiclassical  partition functions),
   similarly to   what was   done  in the previous section
      for the GS and PS  formulations  of the $AdS_5\times S^5$ superstring.

\subsection{Fermionic part of the GS action}
The quadratic fermionic part  of the GS Lagrangian on $AdS_2\times S^2\times T^6$ in conformal gauge
can be written as\footnote{In what follows we consider the case of the
type IIA $AdS_2\times S^2\times T^6$ supergravity background discussed in Section 3.1 of \cite{Sorokin:2011rr}.}
\begin{eqnarray}\label{GS}
 \mathcal{L}_{\rm GS}&=&i\Theta\left(\eta^{ij}-\varepsilon ^{ij}{\Gamma_{11} }\right)e_i{}^{A}\Gamma_{A}
 \,
 \left(\nabla_j+\frac{1}{2}\mathcal{P}_8\gamma \Gamma _{11}\Gamma_{B} e_j{}^{B}\right)\Theta\,,
\end{eqnarray}
where $\Theta$ is a 32 component Majorana spinor, $\gamma=\Gamma^{01}$ and $\mathcal P_8$ is a projector of rank 8, whose presence implies that in $AdS_2\times S^2\times T^6$ the 32--component supersymmetry of the $D=10$ type IIA vacuum is broken down to 8 (see \cite{Sorokin:2011rr} for more details).
Splitting  $\Theta$ as\foot{As in the $AdS_5\times S^5$ case, the labels $1,3$ refer to the $\mathbbm Z_4$-grading of the underlying supercoset $\frac{PSU(1,1|2)}{SO(1,1)\times SO(2)}$.}
$\Theta^1=\frac{1}{2}(1+\Gamma_{11})\Theta$ and $\Theta^3=\frac{1}{2}(1-\Gamma_{11})\Theta$
we get
\begin{eqnarray}\label{GS-conf}
\mathcal{L}_{\rm GS}&=&
-i\,\Theta^1\slashed e_+\nabla_-\Theta^1
-i\,\Theta^3\slashed e_-\nabla_+\Theta^3
-i\,\Theta^1\slashed e_+\mathcal{P}_8\gamma\slashed e_-\Theta^3\,.
\end{eqnarray}
Since the  bosonic vielbeins
satisfy the on-shell equations \eqref{beomads5}, \eqref{virasoroads5}, \eqref{torsionless},
the matrices $\slashed e_\pm = e_\pm{}^A\Gamma_A$ are covariantly (anti)holomorphic and square to zero like in \eqref{gamma5pm}. As in the $AdS_5\times S^5$ case, the kappa--symmetry of the GS formulation manifests itself in the fact that only half of the  32   fermions $\Theta^{1,3}$, projected respectively with $\slashed e_+$ and $\slashed e_-$, appear in the action \eqref{GS-conf}.

It should be noted that the components of $\Theta$ projected with $\mathcal P_8$ and $\mathcal P_{24}=1-\mathcal P_8$ have different geometrical meaning. Eight fermions  $\vartheta=\mathcal P_8\Theta$ parametrize the Grassmann--odd directions of the supercoset $\frac{PSU(1,1|2)}{SO(1,1)\times U(1)}$, while
twenty four fermions $\upsilon=(1-\mathcal P_8)\Theta=\mathcal P_{24}\Theta$ correspond to the 24 broken
target-space  supersymmetries. In the Lagrangian $\vartheta$ and $\upsilon$ couple to each other through the bosonic modes in $T^6$ as can be seen by rewriting \eqref{GS} in the more explicit form (see \cite{Sorokin:2011rr} for details)
\begin{eqnarray}\label{fermLIIAgs}
 \mathcal{L}_{\rm GS}
&=&i\vartheta\left(\eta^{ij}-\varepsilon ^{ij}{\Gamma_{11} }\right)e_i{}^{\underline a}\Gamma_{\underline a}
 \,
 \mathcal D_j\,\vartheta\nonumber\\
 &&{}+i\vartheta \left(\eta^{ij}-\varepsilon ^{ij}\Gamma_{11}\right)\Gamma_{a'}\nabla_j\upsilon \,\partial_iy^{a'}
 +i\upsilon \left(\eta^{ij}-\varepsilon ^{ij}\Gamma_{11}\right)\Gamma_{a'}\nabla_j\vartheta \,\partial_iy^{a'}\nonumber\\
&&{}+\frac{i}{2}\vartheta\left(\eta^{ij}-\varepsilon ^{ij}{\Gamma_{11} }\right)e_i{}^{\underline a}\Gamma_{\underline a}\,\mathcal{P}_8\gamma \Gamma _{11}\Gamma_{a'}
\upsilon\,\partial_jy^{a'}\\
&&{}+i\upsilon\left( \eta^{ij}-\varepsilon ^{ij}{\Gamma_{11} }\right)(e_i{}^{\underline a}\Gamma_{\underline a}+e_i{}^{ a'}\Gamma_{a'})
 \nabla_j\upsilon \cr
 && \
 +\frac{i}{2 }\,\upsilon\left(
 \eta^{ij}-\varepsilon ^{ij}{\Gamma_{11} }
 \right)\Gamma_{a'}\mathcal{P}_8\gamma \Gamma _{11}\Gamma_{b'}\upsilon\ \partial_i y^{a'}\partial_j y^{b'}\,,\nonumber
\end{eqnarray}
where
\begin{equation}\label{defsforGSIIA}
\nabla_j=\partial _j-\frac{1}{4}\,\Gamma _{\underline{ab}}\,\omega_j{}^{\underline{ab}}(x)\,,\qquad \mathcal D_j=\nabla_j+\frac{1}{2 }\mathcal{P}_8\gamma \Gamma _{11}\Gamma_{\underline b} e_j{}^{\underline b}
\end{equation}
and $\omega _i{}^{\underline{ab}}(x)$ and $e_i{}^{\underline{a}}(x)$  ($\underline a,\underline b=0,1,2,3$) are the worldsheet pull--backs of the spin connection and the local frame in $AdS_2\times S^2$.
Note that the $AdS_2\times S^2$ and $T^6$ sectors are coupled
  via the couplings between $\vartheta$ and \nolinebreak $\upsilon$.


\subsection{Fermionic part of the hybrid model action}\label{fhybrid}

The hybrid model in $AdS_2\times S^2\times T^6$ consists of a supercoset sigma--model on $\frac{PSU(1,1|2)}{SO(1,1)\times U(1)}$ (similar to that in the pure--spinor formulation of the $AdS_5\times S^5$ superstring) and an RNS--like string model on $T^6$ \cite{Berkovits:1999zq} . In contrast to the GS superstring, in the hybrid model action the $\frac{PSU(1,1|2)}{SO(1,1)\times U(1)}$ sector containing $AdS_2\times S^2$ is completely decoupled from the $T^6$ sector.

In the conventions of \cite{Sorokin:2011rr}, the $AdS_2\times S^2$ supercoset part of the hybrid model restricted to the second order in the eight coset fermions $\vartheta=\mathcal P_8\Theta$ has the following form
\begin{eqnarray}\label{fermLIIA}
 \mathcal{L}_{\rm H}
&=&i\vartheta\left(\eta^{ij}-\varepsilon ^{ij}{\Gamma_{11} }\right)e_i{}^{\underline a}\Gamma_{\underline a}
 \,
 \mathcal D_j\,\vartheta\\
 &&+\, 2i \eta^{ij}\,\mathcal D_i\vartheta \, \gamma\Gamma_{11}\,\mathcal D_j\,\vartheta\,,\nonumber
\end{eqnarray}
where there is no need to include the projector $\mathcal{P}_8$ in $\mathcal D_j$ in \eqref{defsforGSIIA}  as
 it commutes with $\Gamma_{\underline a}$.

In the $T^6$ sector the hybrid model contains  six RNS--like 2--component worldsheet spinors $\Psi^{a'}$
\begin{equation}\label{Psi}
L_{\Psi}=\frac{i}{2}\Psi^{a'}\gamma^i\partial_i\Psi^{a'}\,,
\end{equation}
where $\gamma^i$ is a 2--dimensional gamma--matrix. The model also includes a ghost sector which consists of a chiral and an anti--chiral boson. The whole construction possesses $N=(2,2)$ worldsheet superconformal symmetry for which the total central charge is zero (see \cite{Berkovits:1999zq} for more details). As we have already mentioned, in this model the $AdS_2\times S^2$ supercoset sector is  decoupled from the $T^6$ one
(apart from indirect relation via the  $N=(2,2)$ super--Virasoro constraints).

The first line in \eqref{fermLIIA} is the same as the supercoset GS fermion term in the first line of
 eq.\eqref{fermLIIAgs}  and the second line is similar to the the second--derivative  fermion term in the PS action.
Adding a total derivative term, the Lagrangian \eqref{fermLIIA} can be rewritten in the following form
\begin{eqnarray}\label{fermLIIA1}
 \mathcal{L}_{\rm H}
&=&2i\,\nabla_i\vartheta\, \gamma\Gamma_{11}(\eta^{ij}-\varepsilon^{ij}\Gamma_{11})\,(\nabla_j\vartheta+\frac{1}{2}\gamma \Gamma _{11}\Gamma_{\underline b} e_j{}^{\underline b}\vartheta)\,.
\end{eqnarray}
Introducing  $\vartheta^1=\frac{1}{2}(1+\Gamma_{11})\vartheta$ and $\vartheta^3=\frac{1}{2}(1-\Gamma_{11})\vartheta$
we get
\begin{equation}\label{pm}
{\mathcal L}_{\rm H}=-4i\nabla_+\vartheta^1\Gamma_{01}\nabla_-\vartheta^3
+i\vartheta^1 e_-{}^{\underline a}\Gamma_{\underline a}\nabla_+\vartheta^1
+i\vartheta^3 e_+{}^{\underline a}\Gamma_{\underline a}\nabla_-\vartheta^3\,,
\end{equation}
where due to the bosonic equations of motion \eqref{beomads5} and the Virasoro constraints \eqref{virasoroads5}
we have
 \begin{equation}\label{Gammapm}
 \nabla_+(e_-{}^{\underline a}\Gamma_{\underline a})=0=\nabla_-(e_+{}^{\underline a}\Gamma_{\underline a}) \ , \qquad \qquad
(e_\pm{}^{\underline a}\Gamma_{\underline a})^2=-(e_\pm{}^{a'}\Gamma_{a'})^2\,.
\end{equation}
Here $e_\pm{}^{a'}(y)=\partial_\pm y^{a'}$ are pullbacks of the $T^6$ vielbeins.

Comparing eq. \eqref{pm} with (the supercoset part of) the GS action \eqref{GS-conf} we see (as in the pure--spinor formulation of Section 2) that their first--derivative terms are similar up to the interchange of $\vartheta^1$ and $\vartheta^3$, while the mass--like term of the GS action in the hybrid model  gets replaced by the second--derivative term.

The difference of the hybrid model from the PS superstring
 is that in the hybrid model in addition to the supercoset sector we have the $T^6$ sector which includes RNS--like fermions. So we should distinguish two cases: (i) When the classical
   string moves entirely in $AdS_2\times S^2$ (i.e. $e_\pm{}^{a'}(y)=\partial_\pm y^{a'}=0$) and
   (ii)   when the string also moves in $T^6$.

In the first case $(e_\pm{}^{a'}\Gamma_{a'})^2$   vanishes and the problem reduces to the analysis of the supercoset part of the hybrid model. This can be done in the same way as for the PS superstring in Section 2.2. One then concludes that around a generic classical string solution in $AdS_2\times S^2$ the hybrid model has 2 pairs of supercoset fermionic modes that produce the same functional determinant as the GS supercoset fermions.

However, the matching of the contributions 
of the other fermionic modes and the ghosts in the two models turns out to be less straightforward than in the
GS vs  pure--spinor case. Already in  flat space, to relate the hybrid  model to the
 GS model  in the light--cone gauge it is necessary to bosonise 2 of the 6 extra pairs of supercoset hybrid--model fermions into an additional pair of chiral and anti--chiral scalars \cite{Berkovits:1996bf}.
 A similar procedure is expected to apply  in a curved background as well.\footnote{We thank Nathan Berkovits for
  comments on this issue.}

  Assuming this is  the case, the comparison of the contributions into the 1--loop partition functions
  in the two formulations goes as follows.
The bosonic sectors of the two models coincide.
In the GS formulation 2 pairs of supercoset fermionic modes $\vartheta$ contribute to the partition function
with  $\det \nabla_{\rm GS}$. The 6 pairs of non--coset fermions $\upsilon$  give   $(\det \partial^2)^6$ and the conformal
ghosts  give $(\det \partial^2)^2$,  in total  $(\det \partial^2)^6(\det \partial^2)^2=(\det \partial^2)^8$.
In the hybrid model the 2 pairs of the supercoset fermionic modes produce the same determinant as
in  the GS case, i.e. $\det \nabla_{\rm GS}$. The one chiral and one anti--chiral scalar ghost \cite{Berkovits:1999zq} give  $(\det \partial^2)^{-1}$. Two of the extra 6 coset fermions should be bosonised into an additional pair of chiral and anti--chiral scalars which  give  an extra $(\det \partial^2)^{-1}$, while the remaining 4 pairs of coset fermions and the RNS--like fermions $\Psi^{a'}$     give  $(\det \partial^2)^{10}$.  Putting all
  the contributions together we find  that the resulting hybrid model
  partition function coincides with the one of the GS  model.

The case of  generic motion of the string in $AdS_2\times S^2\times T^6$ is more complicated
 (since, in particular, the Virasoro constraints \eqref{Gammapm} relate the $AdS_2\times S^2$ and $T^6$ sectors) and it should be treated separately.
 In what follows we shall demonstrate that the matching of the ``massive'' fermionic modes take place at least  in
  two simple examples --   for  a  BMN-type   geodesic running along $S^2$ and $T^6$
  and for a classical string wrapping a circle in $T^6$.
  We will also compare the GS and hybrid model  fermionic actions in
  the  background of world sheet instantons.

\subsection{Examples of equivalence 
}

\subsubsection{Expansion near BMN geodesic} \label{BMNlimit}

BMN  limits for $AdS_p\times S^q$ backgrounds
and their supersymmetries were considered, e.g.,   in \cite{Cvetic:2002si}. In $AdS_2\times S^2\times T^6$ there are different ways to take the BMN limit. One way is to  consider
a geodesic running along $S^2$ but a more general possibility is to take a geodesic
 along both $S^2$ and $T^6$.
These  cases  were considered in detail in \cite{Sorokin:2011rr} .
Let us choose the metric of $AdS_2\times S^2\times T^6$ in the form (we set the radii to 1)
\begin{equation}\label{met}
ds^2=-\cosh^2\rho\,dt^2+d\rho^2+d\theta^2+\sin^2\theta\,d\varphi^2 +dy_{a'}dy^{a'}\,.
\end{equation}
If the geodesic representing
the center of mass of the string runs along a ``diagonal'' direction in
the $S^1 \times S^1$  torus formed by
the equator $S^1 \subset S^2$ (with coordinate $\varphi$)
  and one of the
$S^1 \subset T^6$ directions, \emph{e.g.},  $y^4$,  it
 can be  parameterized by the ``rotated'' coordinate
  $\varphi'$  as
\begin{eqnarray}\label{primebasis}
t= m \tau \ , \ \ \ \varphi'= m \tau \ , \ \ \ \ \ \ \ \ \
\varphi'=\cos\alpha\ \varphi+\sin\alpha\ {y^4}\,,\ \ \ \ \ \ y^{4'}=-\sin\alpha\ \varphi+\cos\alpha\ y^4\,,
\end{eqnarray}
where $\alpha$  is related to the ratio  of string angular  momenta
along the two circles.
$\alpha=0$ is the  case  when the geodesic runs along $\varphi$ and of $S^2$   and
$\alpha= \frac{\pi}{2}$
corresponds  to  a geodesic  running  along $S^1 \subset T^6$.
The resulting quadratic Lagrangian of the bosonic fluctuations is

\begin{equation}\label{BMNb1}
\mathcal L_{\rm bose}=\frac 12\partial _i\tilde\rho \partial^i\tilde\rho+\frac 12 \partial_i\tilde\theta \partial^i\tilde\theta  +
\frac{m^2}{2}\tilde\rho^2+\frac 12m^2\cos^2\alpha\ \tilde\theta^2
+\frac 12 \partial_iy^{a'} \partial^iy_{a'}\,,
\end{equation}
i.e.  $\tilde\rho$ has mass $m$ and $\tilde\theta$ has mass $m\cos\alpha$.

The GS  Lagrangian for the fermionic fluctuations around the
generic BMN solution has the following form in the light--cone kappa--symmetry gauge $\Gamma^+\Theta=0$
\begin{eqnarray}\label{GSBMNF}
\mathcal L_{\rm GS}&=&
-i\Theta^1\Gamma^{-}\partial_-\Theta^1
-i\Theta^3\Gamma^{-}\partial_+\Theta^3
+im(1+\cos\alpha)\,\tilde\vartheta^1\Gamma^{-}\Gamma^r\tilde\vartheta^3\nonumber\\
&&+im(1-\cos\alpha)\,\tilde\upsilon^1\Gamma^{-}\Gamma^r\tilde\upsilon^3,
\end{eqnarray}
where  the labels 1 and 3 refer  again to the $\mathbbm Z_4$-grading (i.e. the chirality)
 and $\Gamma^r$  corresponds  to the radial direction of $AdS_2$.
The spinors $\tilde\vartheta$ and $\tilde\upsilon$ are four--component fermions among $\Theta^{1,3}$ which carry two physical degrees of freedom each,
 having, respectively, mass $\frac{m}2(1+\cos\alpha)$ and $\frac{m}2(1-\cos\alpha)$ (see \cite{Sorokin:2011rr} for more details). The  sums of the mass squared  terms   for the bosonic  and fermionic modes  match  in agreement with UV finiteness   of the model.

\medskip

Let us now turn to the hybrid model and start with the $\alpha=0$ case  (i.e. when the geodesic  runs inside $S^2$).
The corresponding BMN  limit of the hybrid model was considered in \cite{Berkovits:2002zv}.
Here  there is  no kappa--symmetry, so we cannot impose
 $\Gamma^+\vartheta=0$. Let us denote  half of    $\vartheta $
  which satisfies  $\Gamma^+\vartheta=0$ by 
  $X^1=\frac{1}{2}(1+\Gamma_{11})X^1$, $X^3=\frac{1}{2}(1-\Gamma_{11})X^3$
  and  the other half 
  by $Y^1$, $Y^3$, i.e.
  \be\label{rescale}
(X^1,X^3)= \Gamma^+\Gamma^-\vartheta\,,\qquad (Y^1,Y^3)=
\Gamma^-\Gamma^+\vartheta\,.
\ee
 To  the  leading order we consider   we may
  replace $e_i{}^{\underline a}\Gamma_{\underline a}\rightarrow-2m\,\delta_i^0\,\Gamma^-$.
Then  from \eqref{pm}   we find
\bee\label{BMN}
\mathcal{L}_{\rm H}
&=&
-2im\,X^1\Gamma^-\partial_+X^1
-2im\,X^3\Gamma^-\partial_-X^3
+4i\,\partial_+X^1\Gamma^-\Gamma^r\partial_-X^3\nonumber\\
&&+4i  \,\partial_+Y^1\Gamma^+\Gamma^r\partial_-Y^3\,,
\eee
where $\Gamma^r$  corresponds again to the spatial direction of $AdS_2$.
Written in  first order form for  the $X$--fermions  \rf{BMN} reads
\bee\label{BMN1}
\mathcal{L}_{\rm H}
&=&
-i\psi^1\Gamma^-\Gamma^r\psi^3
+2i\psi^1\Gamma^-\Gamma^r\partial_-X^3
+2i\,\partial_+X^1\Gamma^-\Gamma^r\psi^3
\nonumber\\
&&{}
-2im\,X^1\Gamma^-\partial_+X^1
-2im\,X^3\Gamma^-\partial_-X^3+4i\,\partial_+Y^1\Gamma^+\Gamma^r\partial_-Y^3\,.
\eee
The redefinitions
\begin{equation}\label{BMN2}
X^1=\frac{1}{2\sqrt{m}}(\tilde X^1-\Gamma^r\tilde\vartheta^3)\,,\quad X^3=\frac{1}{2\sqrt{m}}(\tilde X^3+\Gamma^r\tilde\vartheta^1)
\,,\quad\psi^1=\sqrt{m}\,\tilde\vartheta^1
\,,\quad\psi^3=\sqrt{m}\,\tilde\vartheta^3\,
\end{equation}
lead to
\bee\label{BMN3}
\mathcal{L}_{\rm H}
&=&
\frac{i}{2}\,\tilde\vartheta^1\Gamma^-\partial_-\tilde\vartheta^1
+\frac{i}{2}\,\tilde\vartheta^3\Gamma^-\partial_+\tilde\vartheta^3
-im\tilde\vartheta^1\Gamma^-\Gamma^r\tilde\vartheta^3
-\frac{i}{2}\,\tilde X^1\Gamma^-\partial_+\tilde X^1
-\frac{i}{2}\,\tilde X^3\Gamma^-\partial_-\tilde X^3
\nonumber\\
&&{}
+4i\,\partial_+Y^1\Gamma^+\Gamma^r\partial_-Y^3\,.
\eee
We read off that  $\tilde\vartheta^{1,3}$ are two physical fermionic modes of mass $m$.
Since $\Gamma^+\tilde\vartheta=0$ these match exactly the massive modes of the GS string in \rf{GSBMNF}.

 In the case of  more general   geodesic along $S^2$ and $T^6$
  the action \eqref{fermLIIA1} reduces to
\bee\label{BMNa}
 {\mathcal L}_{H}&=&
-i(1+\cos\alpha)m\,X^1\Gamma^-\partial_+X^1
-i(1+\cos\alpha)m\,X^3\Gamma^-\partial_-X^3\no\\ &&
+4i\,\partial_+X^1\Gamma^-\Gamma^r\partial_-X^3
-i(1-\cos\alpha)m\,Y^1\Gamma^+\partial_+Y^1\no\\ &&
-i(1-\cos\alpha)m\,Y^3\Gamma^+\partial_-Y^3
+4i\,\partial_+Y^1\Gamma^+\Gamma^r\partial_-Y^3\,,
\eee
where 
$\Gamma^{\pm}=\frac{1}{2}\,(\Gamma^0\pm\Gamma^3)$,
with the index $3$ denoting the $\varphi$ direction of $S^2$.
The analysis of the spectrum follows the same lines as in eqs. \eqref{BMN1}--\eqref{BMN3}.
Passing to the first--order  form   
we get
\bee\label{BMN4}
\mathcal{L}_{\rm H}
&=&
\frac{i}{2}\Big[  \,\tilde\vartheta^1\Gamma^-\partial_-\tilde\vartheta^1
+\,\tilde\vartheta^3\Gamma^-\partial_+\tilde\vartheta^3
-i(1+\cos\alpha)m\, \tilde\vartheta^1\Gamma^-\Gamma^r\tilde\vartheta^3
\nonumber\\
&&{}
\ \ +\,\tilde\upsilon^1\Gamma^+\partial_-\tilde\upsilon^1
+,\tilde\upsilon^3\Gamma^+\partial_+\tilde\upsilon^3
-i({1-\cos\alpha})m\, \tilde\upsilon^1\Gamma^+\Gamma^r\tilde\upsilon^3\\
&&\ \ {}-\tilde X^1\Gamma^-\partial_+\tilde X^1
-\,\tilde X^3\Gamma^-\partial_-\tilde X^3-\,\tilde Y^1\Gamma^+\partial_+\tilde Y^1
- \tilde Y^3\Gamma^+\partial_-\tilde Y^3\Big]\,.\nonumber
\eee
Thus  two pairs of fermions $\tilde\vartheta^{1,3}$ have the mass $\frac{m}{2}(1+\cos\alpha)$ and the two fermions $\tilde\upsilon^{1,3}$ have the mass $\frac{m}{2}(1-\cos\alpha)$. These masses agree with those of the corresponding BMN limit of the GS superstring in
 \rf{GSBMNF}, although all the GS fermions satisfy the condition $\Gamma^+\Theta=0$,
 while in the hybrid case $\vartheta$ and $\upsilon$ are projected with $\Gamma^\pm$, respectively.

As in the pure--spinor model case,
the number of massless fermions of the hybrid model   do not appear to match that of the GS model,
\if{}
In the BMN limit under consideration the GS superstring has four massless fermionic modes, while the hybrid model has
six massless RNS fermions $\Psi^{a'}$ \eqref{Psi} and in addition it has massless ghost--like fermions $\tilde X^{1,3}$ and $\tilde Y^{1,3}$.
\fi
but as was  already discussed in Section \ref{fhybrid}, the contribution of  extra  massless fermions of the hybrid model, upon the bosonization of two of them, should be compensated by
the chiral boson ghosts so that   at the end  there is a match  with the GS formulation.

\subsubsection{String wrapping an $S^1\subset T^6$}

Next, let us now consider a simple  solution in which the string wraps an $S^1$ circle of $T^6$
(we shall label it by    index 9)
\begin{equation}\label{s1}
t=\tau, \qquad y^9=\sigma,\qquad \theta=\frac\pi 2,\qquad \rho=\varphi=y^{4,5,6,7,8}=0.
\end{equation}
The corresponding induced worldsheet metric  is flat.
The Lagrangian of bosonic fluctuations is
\bee\label{s1b}
\mathcal L_{\rm bose}=-\frac{1}{2}(\partial\tilde t)^2 + \frac{1}{2}\big[(\partial\tilde\rho)^2+\tilde\rho^2\big]
+\frac{1}{2}(\pa\tilde\theta)^2+\frac{1}{2}(\pa\tilde \varphi)^2+\frac{1}{2}(\pa \tilde y^{a'})^2\,.
\eee
The quadratic fermionic part of  the GS string Lagrangian  \rf{GS} here  simplifies to
\begin{eqnarray}\label{GSs1}
\mathcal L_{\rm GS}&=&i\Theta\left(1-\Gamma_{09}\Gamma_{11}\right)\slash \!\!\!\partial\Theta
 \,
-\frac{i}{2 }\Theta\left(1-\Gamma_{09}\Gamma_{11}\right) \mathcal{P}_8\Gamma_{01} \Gamma _{11}\Theta\\
&&-\frac{i}{2 }\Theta \left(1-\Gamma_{09}\Gamma_{11}\right)\Gamma_9\mathcal{P}_8\Gamma_{01} \Gamma _{11}\Gamma_9\Theta\,,\nonumber
\end{eqnarray}
where  $\slash \!\!\!\partial=-\Gamma_0\partial_\tau+\Gamma_9\partial_\sigma$.
Introducing  $\Theta_{\pm}=\frac{1}{2}(1\pm\Gamma_{09}\Gamma_{11})\Theta $ we get
\begin{eqnarray}\label{GSs121}
\mathcal L_{\rm GS}&=&2i\Theta_-\slash \!\!\!\partial\Theta_-
 \,
- {2i}{ }\Theta_-\mathcal{P}_8\Gamma_{01} \Gamma _{11}\Theta_-\,
 \,
- {i}{ }\Theta_- \mathcal{P}_8\Gamma_{01} \Gamma _{11}\Theta_++ {i}{ }\Theta_- \mathcal{P}_8\Gamma_{01}\Gamma_{11}\Theta_+\nonumber\\
&=&2i\Theta_-\Gamma_0(-\partial_\tau+\Gamma_{11}\partial_\sigma)\Theta_-
 \,
- {2i}{ }\Theta_-\mathcal{P}_8\Gamma_{01} \Gamma _{11}\Theta_-\,.
\end{eqnarray}
The fermions $\Theta_+$  drop out of the Lagrangian  which is again a manifestation of the $\kappa$--symmetry.
We are thus left with 16 fermions $\Theta_-$. In \eqref{GSs121} their mass term contains the projector $\mathcal P_8$;
 hence,  only $\mathcal P_8\Theta_-$ fermionic modes are massive. To compute the number of their components we should find the rank of the projector $\mathcal P_8(1-\Gamma)$ where $\Gamma= \Gamma_{09}\Gamma_{11}$.  Since $\mathcal P_8\Gamma_9\mathcal P_8=0$, we can rewrite the projector as follows
\begin{equation}\label{P8G}
\mathcal P_8(1-\Gamma)=\mathcal P_8-\mathcal P_8\Gamma(\mathcal P_{24}+\mathcal P_8)=\mathcal P_8-\mathcal P_8\Gamma\mathcal P_{24}\,,
\end{equation}
where $\mathcal P_{24}$ is the projector complementary to $\mathcal P_8$.
The projector complementary to \eqref{P8G} has the form
\begin{equation}\label{compli}
1-\mathcal P_8(1-\Gamma)=\mathcal P_{24}+\mathcal P_8\Gamma\mathcal P_{24}=(1+\mathcal P_8\Gamma)\mathcal P_{24}\,.
\end{equation}
The matrix $1+\mathcal P_8\Gamma$ is invertible, its inverse being $1-\mathcal P_8\Gamma$. Hence, the projector \eqref{compli} has rank 24 and the projector \eqref{P8G} has rank 8. We thus conclude that $\mathcal P_8\Theta_-$ is an 8--component spinor which carries 4  independent physical fermionic modes.

 We should now split the kinetic term of eq. \eqref{GSs121} into two parts, one for the massive eight--component  fermions $\tilde \vartheta=\mathcal P_8\Theta_-$ and another one for the massless 8--component fermions $\psi=(1+\mathcal P_8\Gamma)\mathcal P_{24}\Theta_{-}$ which complement $\tilde\vartheta$ to the 16--component spinor $\frac{1}{2}(1-\Gamma)\Theta$. To this end we insert into the kinetic term the unit matrix $1=\mathcal P_8(1-\Gamma)+(1+\mathcal P_8\Gamma)\mathcal P_{24}$
 \begin{eqnarray}\label{kinetic}
 &2i\Theta_-\Gamma_0(-\partial_\tau+\Gamma_{11}\partial_\sigma)\Theta_-=2i\Theta_-(\mathcal P_8(1-\Gamma)+(1+\mathcal P_8\Gamma)\mathcal P_{24})\Gamma_0(-\partial_\tau+\Gamma_{11}\partial_\sigma)\Theta_-&\nonumber\\
& =4i\tilde\vartheta\Gamma_0(-\partial_\tau+\Gamma_{11}\partial_\sigma)\tilde\vartheta
+2i\psi\Gamma_0(-\partial_\tau+\Gamma_{11}\partial_\sigma)\psi\,.&
 \end{eqnarray}
We  end up with the Lagrangian
\begin{equation}\label{s1fl}
\mathcal L_{\rm GS}=4i\tilde\vartheta\Gamma_0(-\partial_\tau+\Gamma_{11}\partial_\sigma)\tilde\vartheta+2i\tilde\vartheta\Gamma_{01}\Gamma_{11}\tilde\vartheta
+2i\psi\Gamma_0(-\partial_\tau+\Gamma_{11}\partial_\sigma)\psi\ .
\end{equation}
Thus $\tilde\vartheta$   describes
 4 physical fermionic modes of  mass $\frac{1}{2}$, so that the  boson-fermion mass--squared  sum rule is again satisfed.

\medskip

The fermionic  part of the hybrid model   Lagrangian
  \eqref{fermLIIA1}
 takes the form
\begin{eqnarray}\label{fermLIIAs1}
 \mathcal{L}_{\rm H}
&=&i\vartheta \Gamma_0(\partial_\tau+\Gamma_{11}\partial_\sigma)\vartheta -
2i \,\partial_i\vartheta\Gamma_{01}\Gamma_{11}\,\partial^i\vartheta \ .
\end{eqnarray}
In terms of   $\vartheta_{\pm}=\frac{1}{2}(1\pm\Gamma_{01}\Gamma_{11})\vartheta$ we  get
\begin{eqnarray}\label{fermLIIAs11}
 \mathcal{L}_{\rm H}
&=&-i\vartheta_+ \slash\!\!\!\partial\vartheta_+ -
2i \,\partial_i\vartheta_+\partial^i\vartheta_+-i\vartheta_- \slash\!\!\!\partial\vartheta_- +
2i \,\partial_i\vartheta_-\,\partial^i\vartheta_-  \no \\
&=&-i\vartheta_+ \slash\!\!\!\partial\vartheta_+ +
2i \, \slash\!\!\!\partial\vartheta_+ \slash\!\!\!\partial\vartheta_+-i\vartheta_- \slash\!\!\!\partial\vartheta_- -
2i \, \slash\!\!\!\partial\vartheta_-\,\slash\!\!\!\partial\vartheta_-\,,
\end{eqnarray}
where now $\slash\!\!\!\partial=-\Gamma_0(\partial_\tau+\Gamma_{11}\partial_\sigma)$.
We can now pass to the first order form
 by introducing the Lagrange multipliers \nolinebreak $\psi_\pm$
\begin{eqnarray}\label{fermLIIAs122}
 \mathcal{L}_{\rm H}
&=&-i\vartheta_+ \slash\!\!\!\partial\vartheta_+ +
2i\, \psi_+\slash\!\!\!\partial\vartheta_+-\frac{i}{2 }\psi_+\psi_+-i\vartheta_- \slash\!\!\!\partial\vartheta_- -
2i\, \psi_-\,\slash\!\!\!\partial\vartheta_-\,+\frac{i}{2 }\psi_-\psi_-\nonumber\\
&=&
i\tilde\vartheta_+ \slash\!\!\!\partial\tilde\vartheta_+ -
i\,\tilde \psi_+ \slash\!\!\!\partial\tilde\psi_+-\frac{i}{2 }\tilde\vartheta_+\tilde\vartheta_++i\tilde\vartheta_- \slash\!\!\!\partial\tilde\vartheta_- -
i\, \tilde\psi_-\,\slash\!\!\!\partial\tilde\psi_-\,+\frac{i}{2 }\tilde\vartheta_-\tilde\vartheta_-\,,
\end{eqnarray}
where $\tilde\vartheta_\pm=\psi_\pm$ and $\tilde\psi_\pm=\vartheta_\pm\mp\psi_\pm$. We thus
get  4 fermionic modes $\tilde\vartheta_{\pm}$ of  mass $\frac{1}{2 }$ and 4 massless modes
 $\tilde\psi_\pm$ plus 6 massless fermions $\Psi^{a'}$ \eqref{Psi}.
  The spectrum of  massive fermions is again the same as for the GS superstring.

\subsection{Expansion near $S^2$ and $T^2$ worldsheet instantons  
}

Let us now discuss  a less trivial example of a classical solution
-- worldsheet instantons which exist in the Wick--rotated superstring theory in this background, as well as in $AdS_4\times CP^3$ \cite{Cagnazzo:2009zh}\footnote{The Wick
rotation amounts to
replacing the Minkowski signature metric with the Euclidean one,
 $\varepsilon^{ij}$  with
$-i\,\varepsilon^{ij}$ and taking into account that the fermions
$\Theta$ become complex spinors, since there are no Majorana spinors
in ten-dimensional Euclidean space.  The complex conjugate
spinors do not appear in the Wick rotated action  so that the
number of the fermionic degrees of freedom formally remains the same
as before the Wick rotation. Note also that the Euclidean $\gamma$  matrix
is defined as $\gamma=i\Gamma^{0'}\,\Gamma^1$,
where $\Gamma^{0'}$ is the Wick rotated $\Gamma^0$.   Thus  $\gamma^2=1$
as in the case of Minkowski signature.}.
There are two types of instantons. One is when the string wraps $S^2$. This solution is a Wick rotated counterpart of the $AdS_2$--filling string solution considered in Section \ref{infads2}.
For  a (single) worldsheet instanton wrapping $S^2$  the   geometry is that of the two--sphere,
while the string coordinates along $AdS_2\times T^6$ are constant.
For the description of this solution  \cite{Polyakov:1975yp}
 it is convenient to introduce complex coordinates
both in the worldsheet   and in the target space. 
Written in  terms of a complex coordinate $\zeta$ on $S^2$
\begin{equation}
\zeta=\tan\frac{\theta}{2}\,e^{i\varphi}\,,\ \ \ \ \ \ \ \ \
ds^2=d\theta^2+\sin^2\theta\,d\varphi^2\,,
\end{equation}
 the   conformal  gauge   action  on  $S^2$ reads
\be
S_E=\int d^2z\,\frac{|\partial\zeta|^2+|\bar\partial\zeta|^2}{(1+|\zeta|^2)^2}\,.
\ee
This  action has a local minimum  if $\bar\partial\zeta=0$
or $\partial\zeta=0$,
\emph{i.e.} the embedding is given by a holomorphic function
$\zeta=\zeta(z)$ for the instanton or by an anti--holomorphic
function $\zeta=\zeta(\bar z)$ for the anti--instanton.

Another  type of worldsheet instanton  is described by a  Euclidean string worldsheet wrapping a $T^2$ in $T^6$.
In this case the worldsheet coordinates can be directly identified with those of the $T^2$ torus, $y^i=\xi^i, \ \xi= (\xi^1,\xi^2)$.

\subsubsection{Green-Schwarz   action}

Let us   consider    the quadratic fermionic part of the GS  action expanded near these  instanton solutions.

\bigskip
\noindent
{ \bf $S^2$ instanton}

\medskip
\noindent
  In the $S^2$ instanton background the Wick rotated
 fermionic part of the GS action \eqref{fermLIIAgs} reduces to
\begin{eqnarray}\label{S2}
 \mathcal{L}_{\rm GS}
&=&i\vartheta\big(\sqrt{-h}h^{ij}+i\varepsilon ^{ij}{\Gamma_{11} }\big)e_i{}^{\hat a}\Gamma_{\hat a}
 \,
 \big(\nabla_j+\frac{1}{2}\mathcal{P}_8\gamma \Gamma _{11}\Gamma_{\hat a} e_j{}^{\hat a}\big)\,\vartheta\nonumber\\
&&{}+i\upsilon\big( \sqrt{-h}h^{ij}+i\varepsilon ^{ij}{\Gamma_{11} }\big)e_i{}^{\hat a}\Gamma_{\hat a}
 \nabla_j\upsilon\,,
\end{eqnarray}
where $e_i{}^{\hat a}$ are the vielbeins of the instanton $S^2$ sphere parametrized by the worldsheet coordinates. To carry out
the analysis of the fermionic modes in a  covariant way, in \eqref{S2} we have introduced the induced metric $h_{ij}=e_i{}^{\hat a}e_j{}^{\hat b}\delta_{\hat a\hat b}$ on the worldsheet instanton sphere $S^2$. For a particular choice of the $S^2$ coordinates $h_{ij}$ can be chosen to be conformally flat.

\def \bea  {\begin{eqnarray}}
\def \eea  { \end{eqnarray}}

 We see that $\vartheta$ and $\upsilon$ decouple from each other and have the following equations of motion
\bea  \label{vars2}
&&h^{ij}e_i{}^{\hat a}\Gamma_{\hat a}
 \big(\nabla_j+\frac{1}{2}\mathcal{P}_8\gamma \Gamma _{11}\Gamma_{\hat a} e_j{}^{\hat a}\big)(1-\Gamma)\vartheta=0\,,
\\
\label{vs2}
&&h^{ij}e_i{}^{\hat a}\Gamma_{\hat a}
\nabla_j(1-\Gamma)\upsilon=0\,,
\eea
where $\Gamma=-\frac{i}{2\sqrt{h}}\varepsilon^{ij}\Gamma_{ij}\Gamma_{11}$ coincides on the bosonic instanton configuration with the kappa--symmetry projector.

The equation \eqref{vs2} is the massless Dirac equation on $S^2$ which does not have non--trivial solutions, hence  $(1-\Gamma)\upsilon=0$. The equation \eqref{vars2} is the `massive'  Dirac equation (with mass 1 in the inverse radius units)
 whose only regular solutions are the $S^2$ Killing spinors satisfying
\begin{equation}\label{KS2}
\big(\nabla_j+\frac{1}{2}\gamma \Gamma _{11}\Gamma_{\hat a} e_j{}^{\hat a}\big)(1-\Gamma)\vartheta=
\big(\nabla_j+\frac{i}{2}\Gamma_{23} \gamma _{7}\Gamma_{\hat a} e_j{}^{\hat a}\big)(1-\Gamma)\vartheta=0\,,
\end{equation}
where
$\gamma_7=i\Gamma_{456789}$
and we have dropped the projector $\mathcal P_8$ since it commutes with $\Gamma_{\hat a}$ $(\hat a=2,3)$ on $S^2$ and $\gamma_7$.
To see that \eqref{KS2} is indeed the Killing spinor equation on $S^2$ let us redefine the matrices $\Gamma_{\hat a}$ as follows
\begin{equation}
(\Gamma_{23}\Gamma_{\hat a}) \quad \Rightarrow \quad \Gamma_{\hat a}
\end{equation}
and split the four--component spinor $(1-\Gamma)\vartheta$ into the eigenvalues of $\gamma_7$ (recall  that $(\gamma_7)^2=1$)
\begin{equation}\label{pmg7}
(1-\Gamma)\vartheta=\vartheta_++\vartheta_-,\qquad \vartheta_{\pm}=\frac{1}{2}(1\pm \gamma_7)(1-\Gamma)\vartheta\,.
\end{equation}
Then \eqref{KS2} takes the form of the conventional Killing spinor equations on $S^2$ for two 2--component spinors $\vartheta_+$ and $\vartheta_-$
\begin{equation}\label{CKS2}
\big(\nabla_j\pm\frac{i}{2}\Gamma_{\hat a} e_j{}^{\hat a}\big)\vartheta_{\pm}=0\,.
\end{equation}
Thus we conclude that the string instanton wrapping $S^2$ has 4 fermionic zero modes associated with the $S^2$ Killing spinors $\vartheta_{\pm}$.

\bigskip

\noindent
{ \bf $T^2$ instanton}

\medskip
\noindent
When the string worldsheet wraps a $T^2$  (with coordinates $y^i$)
in $T^6$, i.e.  $y^i=\xi^i$
the fermionic part \eqref{fermLIIAgs} of the Wick rotated GS action reduces to
\begin{equation}\label{T2}
 \mathcal{L}_{\rm GS}=
i\vartheta \left(1-\Gamma\right)\Gamma^{i}\partial_i\upsilon
 +i\upsilon \left(1-\Gamma\right)\Gamma^{i}\partial_i\vartheta
+i\upsilon\left(1-\Gamma\right)\Gamma^{i}\partial_i\upsilon
 -\frac{i}{2}\,\upsilon\gamma \gamma _{7}(1-\gamma_5)(1-\Gamma)\upsilon\,,
\end{equation}
where $\Gamma=-\frac{i}{2}\varepsilon^{ij}\Gamma_{ij}\Gamma_{11}=-\gamma_5\gamma_{\tilde 5}$ is the kappa--symmetry projector in this background  and $\gamma_{\tilde 5}=\Gamma_{3456}$ is the product of the gamma--matrices with the indices of the $T^6$ directions orthogonal to the instanton worldsheet.

The fermionic equations of motion which follow from \eqref{T2} are
\begin{equation}\label{T21}
{\mathcal P}_{24}\Gamma^{i}\partial_i\left(1-\Gamma\right)\upsilon -\frac{1}{2}\,{\mathcal P}_{24}\gamma \gamma _{7}(1-\gamma_5)(1-\Gamma)\upsilon=- {\mathcal P}_{24}\Gamma^{i}\partial_i\left(1-\Gamma\right)\vartheta\ ,
\end{equation}
or, equivalently,
\bea \label{T211}
{\mathcal P}_{24}\Gamma^{i}\partial_i\left(1-\Gamma\right)\upsilon +\frac{1}{2}\,{\mathcal P}_{24}\gamma \Gamma _{11}(1-\gamma_{\tilde 5})(1-\Gamma)\upsilon&=&- {\mathcal P}_{24}\Gamma^{i}\partial_i\left(1-\Gamma\right)\vartheta\ ,
 \\   \label{T22}
{\mathcal P}_{8}\Gamma^{i}\partial_i\left(1-\Gamma\right)\upsilon&=&0\  .
\eea
These two equations can be combined into the single one
\begin{equation}\label{T23}
\Gamma^{i}\partial_i\left(1-\Gamma\right)\upsilon -\frac{1}{2}\,\gamma \gamma _{7}(1-\gamma_5)(1-\Gamma)\upsilon=- {\mathcal P}_{24}\Gamma^{i}\partial_i\left(1-\Gamma\right)\vartheta\ .
\end{equation}
Hitting this equation with ${\mathcal P}_{8}\Gamma^j\partial_j$ and taking into account \eqref{T22} and the fact that ${\mathcal P}_{8}\Gamma^j{\mathcal P}_{8}=0$ we find that $\vartheta$ should satisfy the `massless' Laplace equation on $T^2$
\begin{equation}\label{thetaT2}
\partial^i\partial_i\left(1-\Gamma\right)\vartheta=0\,.
\end{equation}
The zero modes of the Laplace operator on $T^2$ are constants, hence the $T^2$ string instanton has 4 fermionic constant modes $(1-\Gamma)\vartheta$, and eq. \eqref{T23} reduces to
\begin{equation}\label{T24}
\Gamma^{i}\partial_i\left(1-\Gamma\right)\upsilon -\frac{1}{2}\,\gamma \gamma _{7}(1-\gamma_5)(1-\Gamma)\upsilon=0.
\end{equation}
From this equation we read that the eight\footnote{This number follows from the analysis of \cite{Cagnazzo:2009zh}. See Subsection 4.2 therein.} modes $\frac{1}{2}(1+\gamma_5)(1-\Gamma)\upsilon$ should satisfy the massless Dirac equation
\begin{equation}\label{T223}
\Gamma^{i}\partial_i\left(1-\Gamma\right)(1+\gamma_5)\upsilon=0,
\end{equation}
and hence are constants. We are thus left with
\bea \label{T24-}
&& \Gamma^{i}\partial_i\left(1-\Gamma\right)\upsilon_- -\gamma \gamma _{7}(1-\Gamma)\upsilon_-=0, \qquad \upsilon_-=\frac{1}{2}(1-\gamma_5)\upsilon\ ,
\\  \label{T222}
&& {\mathcal P}_{8}\Gamma^{i}\partial_i\left(1-\Gamma\right)\upsilon_-=0\  ,
\eea
where the last equation  is the consequence of \eqref{T22} and \eqref{T223}.

To analyze eq.  \eqref{T222}, let us use the explicit form of the projectors ${\mathcal P}_{8}$ and ${\mathcal P}_{24}$  which is similar to that in the case of the GS string instanton on $CP^3$ (see eq. (4.42) of \cite{Cagnazzo:2009zh})
\begin{eqnarray}\label{p2p6}
\mathcal P_8=\frac{1}{8}(2+J)=\frac{1}{4}(1+\rho^3\tilde J)(1-\gamma_{\tilde 5})\ , \ \ \ \ \ \
\mathcal P_{24}=\frac{1}{8}(6-J)=\frac{1}{4}\big(3+\gamma_{\tilde 5}-\rho^3\tilde J(1-\gamma_{\tilde 5})\big)\,,
\end{eqnarray}
where (the indices $3,4,5,6$ denoting the directions in $T^6$ orthogonal to the instanton worldsheet)
\bea
\rho_3=-\frac{i}{2}\varepsilon^{ij}\Gamma_{ij}\,,\qquad \gamma_{\tilde 5}=\Gamma_{3456}
\ , \ \ \ \tilde J=-\frac{i}{4}J_{\tilde a\tilde b}\gamma^{\tilde a\tilde b}=-\frac{i}{8}J_{\tilde a\tilde b}\gamma^{\tilde a\tilde b}(1-\gamma_{\tilde 5})\ ,
\quad\tilde J^2=\frac{1}{2}(1-\gamma_{\tilde 5})
\eea
Inserting the explicit form \eqref{p2p6} of $\mathcal P_8$ into \eqref{T222} we have
\begin{eqnarray}\label{T224}
&&{\mathcal P}_{8}\Gamma^{i}\partial_i\left(1-\Gamma\right)\upsilon_-=\frac{1}{4}\Gamma^{i}\partial_i(1-\rho^3\tilde J)(1-\gamma_{\tilde 5})\left(1-\Gamma\right)\upsilon_-\nonumber\\
&& \ \ \ \ =\Gamma^{i}\partial_i{\mathcal P_{24}}\left(1-\Gamma\right)\upsilon_--\frac{1}{2} \Gamma^{i}\partial_i(1+\gamma_{\tilde5})\left(1-\Gamma\right)\upsilon_-=\Gamma^{i}\partial_i{\mathcal P_{24}}\left(1-\Gamma\right)\upsilon_-=0\,,
\end{eqnarray}
where the term $(1+\gamma_{\tilde5})\left(1-\Gamma\right)\upsilon_-$ is zero
 since $\Gamma=-\gamma_5\gamma_{\tilde 5}$ and
$
(1+\gamma_{\tilde5})\left(1-\Gamma\right)(1-\gamma_5)=(1+\gamma_{\tilde5})\left(1+\gamma_5\gamma_{\tilde 5}\right)(1-\gamma_5)=0.
$
From eq. \eqref{T224} it follows that $(1-\Gamma)\upsilon_-$ is  constant and from \eqref{T24-} it follows that it is
actually zero.

We thus conclude that the $T^2$  string instanton has 12 constant fermionic zero modes:  8  modes $(1+\gamma_5)(1-\Gamma)\upsilon$ and 4 modes $(1-\Gamma)\vartheta$.

\subsubsection{Hybrid model  action}

Let us now      discuss the  corresponding  fermionic  term in the   hybrid model action and   compare   it with the
GS    action.

\bigskip

\noindent
{ \bf $S^2$ instanton}
\nopagebreak
\nopagebreak\\
\nopagebreak\\
\nopagebreak
On the $S^2$ instanton  background  the fermionic part of the hybrid model Lagrangian takes the following form
\begin{equation}\label{hybridS2}
 \mathcal{L}_{\rm ferm}
=2i\,\nabla_i\vartheta\, \gamma\Gamma_{11}(\sqrt{-h}h^{ij}+i\varepsilon^{ij}\Gamma_{11})\,(\nabla_j\vartheta+\frac{1}{2}\gamma \Gamma _{11}\Gamma_{\hat b} e_j{}^{\hat b}\vartheta)\,+\frac{i}{2}\bar\psi^{a'}\gamma_{\hat a}e^{\hat ai}\nabla_i\psi^{a'},
\end{equation}
where $e_i{}^{\hat a}(\xi)$ $(\hat a=2,3)$ are the vielbeins of the instanton sphere, $h_{ij}=e_i{}^{\hat a}e_{j\hat a}$, \  $\gamma^{\hat a}$ are the worldsheet gamma--matrices and  $\psi^{a'}(\xi)$ ($a'$=4,5,6,7,8,9) are 6 two--component worldsheet RNS--like fermions on $T^6$.

The equations of motion of $\psi$ are massless Dirac equations
on $S^2$
\begin{equation}\label{ds2}
\slash\!\!\!\!\nabla\psi=0, \qquad \slash\!\!\!\!\nabla=e^{\hat ai}\gamma_{\hat a}\nabla_i
\end{equation}
which do not have non--trivial regular solutions. The equations of motion of $\vartheta$ are
\begin{equation}\label{hybridvs}
\frac{1}{\sqrt{h}}\nabla_i (\sqrt{h}h^{ij}+i\varepsilon^{ij}\Gamma_{11})\,(\nabla_j\vartheta+\frac{1}{2}\gamma \Gamma _{11}\Gamma_{\hat b} e_j{}^{\hat b}\vartheta)=0\,.
\end{equation}
We can  rewrite them as
\begin{eqnarray}\label{1i}
&(\nabla_i\nabla^i+\frac i{2} \Gamma_{23}\Gamma_{11})\vartheta+\frac 1{2}\gamma\Gamma_{11}\slash\!\!\!\!\nabla\vartheta+\frac i{2\sqrt h} \gamma\Gamma_{11}\Gamma_{11}\varepsilon^{ij}\Gamma_j\nabla_i\vartheta&\nonumber\\
&=(\nabla_i\nabla^i+\frac i{2} \Gamma_{23}\Gamma_{11})\vartheta+\frac 1{2}\gamma\Gamma_{11}(1+\Gamma)\slash\!\!\!\!\nabla\vartheta
=0&\ .
\end{eqnarray}
 To arrive at \eqref{1i} we have used that on $S^2$ of unit radius
\begin{equation}\label{i}
\frac{1}{\sqrt{h}}\varepsilon^{ij}\nabla_i\nabla_j=\frac{1}{2}\Gamma_{23}\ ,\qquad \qquad \varepsilon^{ij}\Gamma_j=-\frac12\varepsilon^{jk}\Gamma_{jk}\Gamma_{i}=-\Gamma_{23}\Gamma_i\,.
\end{equation}
$\Gamma=-i\Gamma_{23}\Gamma_{11}$ is the same as in eq. \eqref{vs2} and $\gamma=i\Gamma^{01}$ (with $\gamma^2=1$) is the product of Wick rotated $AdS_2$  Dirac matrices.

To analyze the solutions of \eqref{1i}, let us multiply it by the projectors $\frac 12 (1\pm\Gamma)$.
This gives the equations of motion for $\vartheta_\pm =\frac 12 (1\pm\Gamma)\vartheta$
\begin{eqnarray}\label{3i}
&(\nabla_i\nabla^i-\frac 1{2})\vartheta_++\gamma\Gamma_{11}\slash\!\!\!\!\nabla\vartheta_+=0\,,&
 \\  \label{2i}
&(\nabla_i\nabla^i+\frac 1{2})\vartheta_-=0\,.&
\end{eqnarray}
\if{}
Note that in the light--cone frame $\nabla_{\pm}=i\nabla_2\pm\nabla_3$, $\Gamma_\pm=\frac 12(i\Gamma_2\pm\Gamma_3)$ eqs. \eqref{3i} and \eqref{2i} take the form
\begin{eqnarray}\label{lc}
&-\frac 12({\nabla_-\nabla_++\nabla_+\nabla_-}+1)\vartheta_+-\gamma\Gamma_{11}(\Gamma_{+}\nabla_-+\Gamma_-\nabla_+)\vartheta_+=0\,,&
\nonumber\\
\\
&\frac 12(-{\nabla_-\nabla_+-\nabla_+\nabla_-}+1)\vartheta_-=0\,&\nonumber
\end{eqnarray}
We can further split $\vartheta_{\pm}$ with respect to their $\mathbbm Z_4$--grading, namely $\vartheta^1_\pm=\frac 12(1+\Gamma_{11})\vartheta_{\pm}$, $\vartheta^3_\pm=\frac 12(1-\Gamma_{11})\vartheta_{\pm}$, Then the above equations take the form similar to that discussed in \cite{Aisaka:2012ud} (see eq. (3.16) therein), but in which we have also made additional projections with the kappa--symmetry generator\footnote{Recall that the kappa--symmetry projectors act as $\vartheta_{\pm}=\frac 12 (1\mp i \Gamma_{23}\Gamma_{11})\vartheta$, and we further split the fermions into the eigenvalues singled out by the projectors $\frac 12(1\pm\Gamma_{11})$ and $\frac 12(1\pm i\Gamma_{23})$.}
\begin{eqnarray}\label{lc2}
&-\nabla_-\nabla_+\vartheta^3_++\gamma\Gamma_{+}\nabla_-\vartheta^1_+=0\,,\qquad -\nabla_+\nabla_-\vartheta^1_+-\gamma\Gamma_{11}\Gamma_-\nabla_+\vartheta^3_+=0\,,&
\nonumber\\
\\
&\nabla_+\nabla_-\vartheta^1_-=0\,,\qquad \nabla_-\nabla_+\vartheta^3_-=0\,\,.&\nonumber
\end{eqnarray}
\fi
 Using \eqref{i}    we may  get from  \eqref{3i}
\begin{equation}\label{g-}
\slash\!\!\!\!\nabla\left(\slash\!\!\!\!\nabla-\gamma \Gamma _{11}\right)\vartheta_+=0\,.
\end{equation}
Note that the fermionic operator factorizes into the massless Dirac operator times the massive GS operator
(cf. \eqref{vars2},\eqref{CKS2}).
In the case of the sphere, the only non--trivial solutions of \eqref{g-} are the $S^2$ Killing spinors satisfying  
\begin{equation}\label{killing}
(\slash\!\!\!\!\nabla-\gamma \Gamma _{11})\vartheta_+=0
\end{equation}
as in   \eqref{vars2}, \eqref{CKS2}.
This  gives us 4 fermionic zero modes as in the case of the GS instanton on $S^2$.
Note that from eqs. \eqref{3i} and \eqref{killing} it follows that $\vartheta_+=\frac 12(1+\Gamma)\vartheta$ are eigenfunctions of the Laplace operator with the eigenvalue $-\frac12$
$ (\nabla_i\nabla^i +{\textstyle \frac{1}{2}})\vartheta_+=0\,.$
According to  \eqref{2i},  four $\vartheta_{-}$ are the eigenfunctions of the Laplace operator with the
eigenvalue $-\frac12$. Hence, they can also be associated with the $S^2$ Killing spinors with an
 effective mass $\pm 1$.

Thus we seem to get an additional $8=4\times 2$ fermionic zero modes in  the hybrid model  as compared to the corresponding GS
case.  However, the comparison   of the total  expressions for the two partition functions, that we do not attempt here,
requires a  proper definition of the path integral measure, as was already emphasized   on the
example  of  the infinite string in $AdS_2$  in section \ref{infads2}. In fact, since this case is essentially an analytic continuation of that discussion we expect a match also in this case.



\bigskip
\noindent
{\bf $T^2$ instanton}
\nopagebreak\\
\nopagebreak\\
In this case the fermionic part of the hybrid model Lagrangian takes the following form
\begin{eqnarray}\label{hybridT2}
 \mathcal{L}_{\rm ferm}
&=&2i\,\partial^i\vartheta\, \gamma\Gamma_{11}\partial_j\vartheta+\frac{i}{2}\bar\psi^{a'}\gamma^i\partial_i\psi^{a'}\,.
\end{eqnarray}
{}From eq.\eqref{hybridT2} we see that eight $\vartheta$ should obey the Laplace equation on $T^2$ and twelve $\psi^{a'}$ should satisfy the massless Dirac equation, i.e.
\begin{equation}\label{t2em}
\partial^i\partial_i\vartheta=0, \qquad \qquad \gamma^i\partial_i\psi^{a'}=0\ .
\end{equation}
These equations have only constant solutions on $T^2$, so that we get
 20 zero modes associated with them.
 Though this number   does not match the one
 of the GS string instanton on $T^2$ which has 12 constant fermionic modes,
 the difference should presumably be  accounted  for  by the  contribution of  the ghost sector of the hybrid model.

\section{Concluding remarks}

In this paper we have demonstrated the  semiclassical equivalence  between
the Green--Schwarz and the
 pure--spinor formulations  of the $AdS_5\times S^5$ superstring expanded  near
    generic classical  string solutions,
extending earlier  work  in this direction  \cite{Roiban:2007,Aisaka:2012ud}.
 We have also  studied a similar relation between the $AdS_2\times S^2\times T^6$  GS superstring
 and the corresponding hybrid model.

 We have shown that in the   $AdS_5\times S^5$  pure--spinor  model expanded
  around a classical string solution, half of the fermionic modes
   enter the quadratic fluctuation action only linearly.
   Therefore, they can be integrated out,  contributing a  massless $2d$  d'Alembertian operator
    determinant  to the one--loop partition function.  The
    action for the remaining half of the fermions takes
    a   form which resembles the structure  of the GS superstring action   with the first--derivative terms being the same
     but with the GS ``mass''  term   being replaced by  a second--derivative
     fermionic kinetic  term.
      We have found   that 
    the contribution   of these remaining fermions  to the one--loop pure--spinor partition function is
      the same as the non-trivial ``massive'' determinant of  the GS fermions  times an additional  massless determinant,
      with the latter being   cancelled  against the ghost  sector   contributions.

One  may  expect  a similar equivalence  between the GS  and the pure-spinor formulation  also
for the case of the $AdS_2\times S^2\times T^6$    background.
 In this case, however,   there exists also a simpler version  of a  ``pure--spinor''-like formulation
-- the hybrid model  of \ci{Berkovits:1999zq}  in which the  $AdS_2\times S^2$   and $T^6$ parts
 are  essentially decoupled in the action  (in contrast to what happens   in the GS case \ci{Sorokin:2011rr}).
We   have  found that there is  a similar semiclassical   correspondence
 between the  $AdS_2\times S^2 \times T^6$  GS theory
 and the hybrid model by considering particular cases
 when the  classical  strings move  only
 in $AdS_2\times S^2$  and  two simple cases when  the string also moves in  or wraps around a  circle in $T^6$.
 We  also
  compared the spectra of the fermionic zero modes of   $S^2$ and $T^2$ worldsheet instantons 
  in the  GS and the   hybrid   formulations.
It would be interesting to complete  this  analysis   in order to prove the
semiclassical  equivalence  between these  two   superstring   formulations
    for a generic  motion of the string in $AdS_2\times S^2 \times T^6$.


Another   open problem is to  repeat a similar study in the case of superstrings in $AdS_3\times S^3\times T^4$.
There   exists  three    a priori  different      formulations of
superstring theory in this  background    (which can be    supported, in general,  by a   combination of   R-R and NS-NS  3-form fluxes):
(i) The Green-Schwarz  formulation   based on the $ PSU(1,1|2) \times PSU(1,1|2) / SU(1,1) \times SU(2) $
supercoset   sigma model   with a   particular WZ  term, enlarged with 4 bosonic $T^4$ degrees of freedom and additional 16 fermionic ones (see \cite{Rahmfeld:1998zn,Pesando:1998wm,Babichenko:2009dk,Sundin:2012gc,Cagnazzo:2012se}
 and references there);\footnote{In the absence of R-R flux there is of course also the standard  RNS    formulation based   on   $2d$ supersymmetric $SU(1,1) \times SU(2)$ WZW  model.}
(ii)   the supercoset hybrid   model   with a  ``pure--spinor''-like (second-derivative fermion)
 sigma model  action   \ci{Berkovits:1999zq,Berkovits:1999du}
 ( for   the same supercoset with the same WZ term and  the same number of 16 supersymmetries as in the GS case);
 (iii) the supergroup  hybrid model \cite{Berkovits:1999im}       based on the $ PSU(1,1|2)$
 sigma model (with an  extra  WZ term  in the case of   non-zero   NS-NS 3-form flux)
  in which only 8 out of 16 supersymmetries are manifest.\footnote{For some recent   discussions of this supergroup
   hybrid  model see, e.g.,    \ci{Ashok:2009jw,Gaberdiel:2011vf,Gerigk:2012cq}
    and references there.}
 It would be interesting to check that these  three formulations are indeed  equivalent
 in the semiclassical expansion.


\subsection*{Acknowledgements}
The authors are grateful to I. Bandos, N. Berkovits, J. Gomis, R. Roiban, I. Samsonov,
M. Tonin,  B. Vallilo  and K. Zarembo for useful discussions and comments.
A.A.T. is grateful to R. Roiban   for a collaboration on  a related unpublished work \ci{Roiban:2007}.
Work of A.C., D.S. and A.A.T.
was partially supported by
 the Uni--PD Research Grant CPDA119349. D.S.
 was also supported in part by the MIUR-PRIN contract 2009-KHZKRX.
 A.C. is thankful to Angelo della Riccia Foundation for the grant to visit Nordita.
The work of A.A.T.
was supported by the STFC grant ST/J000353/1 and by the ERC Advanced grant No.290456.
The research of L.W. was supported in part by NSF grants PHY-0555575 and PHY-0906222.

 \def\thesection{}
\def\theequation{A.\arabic{equation}}
\section{Appendix A. Notation and conventions \label{A}}
\setcounter{equation}0
The  flat metric
is $\eta_{AB}=(-,+\cdots,+)$. The $D=10$ gamma matrices $\Gamma^A$ and $\Gamma^{11}$ are real and
$$
\{\Gamma^A,\Gamma^B\}=2\eta^{AB}.
$$
Contracted with the complex charge conjugation matrix $C$ the matrices
$$
C\Gamma^{\hat A},\qquad  C\Gamma^{\hat A\hat B},\quad C\Gamma^{\hat A_1\cdots \hat A_5}\qquad {\rm are ~symmetric}\qquad (\hat A,\hat B,...=0,\ldots,9,11)
$$
and the matrices
$$
C\,,\qquad C\Gamma^{\hat A\hat B\hat C},\quad C\Gamma^{\hat A \hat B\hat C\hat D}\qquad {\rm are ~anti-symmetric}.
$$
Any two spinors are contracted as follows
$$ \theta^\alpha C_{\alpha\beta}\psi^\beta \equiv\theta\psi,\qquad  \theta C\Gamma^{\hat A...}\psi= \theta^\alpha (C\Gamma^{\hat A...})_{\alpha\beta}\psi^\alpha\equiv \theta\Gamma^{\hat A...}\psi.$$
Note that to simplify the notation we  do not   indicate explicitly
the  charge conjugation matrix $C$ in bilinear combinations of the spinors  contracted with $C\Gamma^{\hat A...}$.

Since the matrices $C\Gamma^{\hat A}$ and $C\Gamma^{\hat A\hat B}$ are symmetric,  we have
$$(\Gamma^{\hat A}\vartheta)C\psi=-\vartheta C\Gamma^{\hat A}\psi\equiv-\vartheta \Gamma^{\hat A}\psi\ , \ \ \ \ \ \ \
(\Gamma^{\hat A\hat B}\vartheta)C\psi=-\vartheta C\Gamma^{\hat A\hat B}\psi\equiv -\vartheta \Gamma^{\hat A\hat B}\psi\,.$$

 \def\thesection{}
\def\theequation{B.\arabic{equation}}
\section{Appendix B. Folded spinning string in $R\times S^2\subset AdS_2\times S^2\times T^6$ }\label{B}

To illustrate the semiclassical equivalence of the GS formulation
 and the hybrid model  for  $AdS_2\times S^2\times T^6$
that we  argued for in general
  when the string is moving entirely in $AdS_2\times S^2$
  let us consider the  case  of a folded spinning string  moving in $S^2$.
  The corresponding solution in conformal gauge is
\begin{eqnarray}  && t= \kappa \tau,\ \  \ \theta= \theta(\s), \ \ \
\varphi= w \tau \ ,\\ &&
\sin \theta = \sqrt q\  {\rm sn} ( w \s | q ) \ , \ \ \ \ q=\sin^2 \theta_0= \frac{\kappa^2 }{ w^2}\ ,
\ \ \ \  w= \frac{ 2}{ \pi} {\rm K}(q) \ ,  \label{kp}
\end{eqnarray}
where $t$ is the $AdS_2$ time coordinate, $\theta$ and $\varphi$ are the spherical coordinates of $S^2$,
 and ${\rm K}$ is the elliptic integral.
 Here $ \theta'^2 = \kappa^2 - w^2  \sin^2 \theta$ so that   the induced metric is
\begin{equation}\label{induced}
ds^2=\theta'^2(-d\tau^2+d\sigma^2)\,.
\end{equation}
The non--zero worldsheet pullbacks of the $AdS_2\times S^2\times T^6$ vielbeins are
\begin{equation}\label{spinv}
e_\tau{}^0=\kappa,\qquad e_\sigma{}^2=\theta', \qquad e_\tau{}^3=\omega\,\sin\theta
\end{equation}
and the covariant derivative acting on the fermions is
\begin{equation}\label{nablaspin}
\nabla_i=\partial_i-\frac{1}{4}\omega_i^{\underline{ab}}\Gamma_{\underline{ab}}=(\pa_\tau-\frac 12 w\cos\theta \Gamma_{23}\,,\,\pa_\sigma)\,, \qquad \nabla_\pm=\partial_\pm-\frac 12 w\cos\theta \Gamma_{23}\,.
\end{equation}
In the conformal gauge, the bosonic fluctuations around this solution are described by the following Lagrangian
(see, e.g., \ci{Beccaria:2010zn})
\bee
&&L_{\rm boson}=-\frac 1 2(\partial \tilde t)^2+\frac12[(\partial \tilde \rho)^2+k^2\tilde\rho^2]
+\frac12(\pa \tilde\theta)^2+\frac 12 \sin^2\theta(\pa\tilde\varphi)^2-2w\cos\theta\sin\theta\, \tilde\theta\pa_\tau \tilde\varphi\nn\\  &&
 \ \ \ \ \  \ \ \ \ \ \ \ \ \ \
-\frac 12(1-2\sin^2\theta)w^2\tilde\theta^2+\frac12(\pa\tilde y)^2.
\eee
Defining $\eta =\sin\theta\,\tilde\varphi$ one gets
\bee
&&L_{\rm boson}=-\frac 1 2(\partial \tilde t)^2+\frac12[(\partial \tilde \rho)^2+k^2\tilde\rho^2]+\frac 12[(\nabla \tilde\theta)^2
+w^2\sin^2\theta\, \tilde\theta^2]\nn\\
 &&\ \ \ \ \ \ \ \ \ \ \  \ \ \ +\frac12[(\nabla \eta )^2-\theta'^2 \eta^2]+\frac 12(\pa\tilde y)^2\,,
\eee
where $\nabla \eta=(\pa_\tau \eta - w \cos\theta\,\tilde \theta,\pa_\sigma \eta)$ and $\nabla \tilde \theta=(\pa_\tau \tilde \theta+ w \cos\theta\, \eta,\pa_\sigma \tilde \theta)$.
Thus  we have three ``massive" bosonic modes, one in $AdS_2$ and two  in $S^2$, with the sum of  squares of their masses   being
\be
\sum m_b^2=2\,w^2\sin^2\theta.
\ee

\subsubsection*{Green--Schwarz action}
The  quadratic  fermonic term in the GS action  \eqref{GS-conf}  in this background
takes the following form
\begin{eqnarray}\label{spinning GS}
 \mathcal{L}_{\rm GS}=
-i\,\vartheta^1\slashed e_+\nabla_-\vartheta^1
-i\,\vartheta^3\slashed e_-\nabla_+\vartheta^3
+i\,\vartheta^1\slashed e_+\Gamma_{01}\slashed e_-\vartheta^3-i\,\upsilon^1\slashed e_+\nabla_-\upsilon^1
-i\,\upsilon^3\slashed e_-\nabla_+\upsilon^3\,.
\end{eqnarray}
Making use of the explicit form of the spin connection \eqref{nablaspin},  let us
 perform the following redefinition of $\vartheta^{1,3}$ and $\upsilon^{1,3}$
\begin{equation}\label{13}
\vartheta^{1}=\frac{1}{\sqrt\kappa} e^{\frac {f(\sigma)}2 \Gamma_{23}}\tilde\vartheta^1\,,\quad \vartheta^{3}=\frac{1}{\sqrt\kappa} e^{-\frac {f(\sigma)}2 \Gamma_{23}}\tilde\vartheta^3\,, \quad \upsilon^{1}=\frac{1}{\sqrt\kappa} e^{\frac {f(\sigma)}2 \Gamma_{23}}\tilde\upsilon^1\,,\quad \upsilon^{3}=\frac{1}{\sqrt\kappa} e^{-\frac {f(\sigma)}2 \Gamma_{23}}\tilde\upsilon^3\,,
\end{equation}
where
\begin{equation}\label{f}
f=\arcsin\frac{\omega\sin\theta}\kappa\,, \qquad \partial_\sigma f=\omega \cos\theta\,, \qquad \cos f=\frac{\theta'}\kappa=\frac{\sqrt{k^2-\omega^2\sin^2\theta}}\kappa\,.
\end{equation}
Then, taking into account that
$$
e^{-\frac f2\Gamma_{23}}\slashed e_- e^{\frac f2\Gamma_{23}}=\kappa(\Gamma_0-\Gamma_2)=\kappa\,\Gamma_-\,,\qquad
e^{\frac f2\Gamma_{23}}\slashed e_+e^{-\frac f2\Gamma_{23}}=\kappa(\Gamma_0+\Gamma_2)=\kappa\,\Gamma_+,
$$
we get
\begin{equation}\label{spinning GS11}
 \mathcal{L}_{\rm GS}
=-2i\tilde\vartheta_-^1\Gamma_0\partial_-\tilde\vartheta_-^1-2i\tilde\vartheta_+^3\Gamma_0\partial_+\tilde\vartheta_+^3 -4i\omega\sin\theta\,\tilde\vartheta_-^1\Gamma_{13}\tilde\vartheta_+^3 +
 2i\tilde\upsilon\slash\!\!\!\partial \tilde\upsilon\,,
\end{equation}
where $\tilde\vartheta^1_-=\frac 12(1-\Gamma_{02})\tilde\vartheta^1$ and  $\tilde\vartheta^3_+=\frac 12(1+\Gamma_{02})\tilde\vartheta^3$.
We thus find that two pairs of fermions $\tilde\vartheta$ have the ``mass" $m=\omega\sin\theta$ and 6 pairs of the fermions $\tilde\upsilon$ are massless.

\subsubsection*{Hybrid model action}
Upon performing the redefinition \eqref{13} of $\vartheta^{1,3}$  and splitting them into $\tilde\vartheta^{1,3}_\pm=\frac 12(1\pm \Gamma_{02})\tilde\vartheta^{1,3}$, the fermionic part of the hybrid model  Lagrangian \eqref{pm} takes the form
 \begin{eqnarray}\label{pm12000}
{\mathcal L}_{\rm H}
&=&\frac{4i\cos f}\kappa\,\partial_+\tilde\vartheta_-^1\Gamma_{12}\partial_-\tilde\vartheta_-^3-\frac{4i\cos f}\kappa\,\partial_+\tilde\vartheta_+^1\Gamma_{12}\partial_-\tilde\vartheta_+^3-\frac{4i\sin f}\kappa\,\partial_+\tilde\vartheta_-^1\Gamma_{13}\partial_-\tilde\vartheta_+^3\nonumber\\
&&+\frac{4i\sin f}\kappa\,\partial_+\tilde\vartheta_+^1\Gamma_{13}\partial_-\tilde\vartheta_-^3+2i\,\tilde\vartheta_+^1 \Gamma_0\partial_+\tilde\vartheta_+^1+2i\,\tilde\vartheta_-^3\Gamma_0\partial_-\tilde\vartheta_-^3\,.
\end{eqnarray}
Using that  $\sin f=\frac{\omega\sin\theta}\kappa$  this Lagrangian
can be    written as
\begin{eqnarray}\label{pm12}
{\mathcal L}_{\rm H}&=&-\frac{4i\omega\sin \theta}{\kappa^2}\,(\partial_+\tilde\vartheta_-^1-\cot f\,\Gamma_{23}\partial_+\tilde\vartheta^1_+)\,\Gamma_{13}\,(\partial_-\tilde\vartheta_+^3+\cot f\,\Gamma_{23}\partial_-\tilde\vartheta_-^3)\nonumber\\
&&+\frac{4i}{\omega\sin \theta}\,\partial_+\tilde\vartheta_+^1\Gamma_{13}\partial_-\tilde\vartheta_-^3+2i\,\tilde\vartheta_+^1 \Gamma_0\partial_+\tilde\vartheta_+^1+2i\,\tilde\vartheta_-^3\Gamma_0\partial_-\tilde\vartheta_-^3\,.
\end{eqnarray}
Note that in the first line of \eqref{pm12} the pairs of terms in each  bracket have the same chirality. This is important for performing the integration of $\tilde\vartheta^1_-$ and $\tilde\vartheta^3_+$ which enter the Lagrangian linearly.
Under an appropriate assumption about the path integral measure (cf.  \cite{Schwarz:1992te})
they  contribute to the partition function  only with a massless determinant  factor $(\det \partial_+\partial_-)^4$.

After  the modes $\vartheta^1_-$ and $\vartheta^3_+$ are integrated out 
we are left with the
last line of \eqref{pm12}, which is
a 
 counterpart of the GS Lagrangian \eqref{spinning GS11}.
 By  the same reasoning  as used  for a  generic classical solution in section 3
 one can verify that this part of the Lagrangian \eqref{pm12}
 indeed describes the same number (two) of
 massive fermions with the same mass as in the GS string case
 plus two massless fermions.

\if{}
\bibliography{strings,self-duality}
\bibliographystyle{utphys}
\end{document}
\fi

\providecommand{\href}[2]{#2}\begingroup\raggedright\endgroup

\end{document}